\documentclass[pdflatex,aps,prd,nofootinbib,twocolumn,showpacs,superscriptaddress]{revtex4-1}
\usepackage{amssymb,amsmath,color,hyperref,graphicx,multirow,tabularx,physics}
\usepackage[switch]{lineno}
\def\lsim{\raise0.3ex\hbox{$<$\kern-0.75em\raise-1.1ex\hbox{$\sim$}}}
\def\gsim{\raise0.3ex\hbox{$>$\kern-0.75em\raise-1.1ex\hbox{$\sim$}}}
\newcommand \Mc {{\cal M}}
\newcommand \Oc {{\cal O}}
\newcommand{\hmu}{\hat{\mu}}
\newcommand{\subl}{\text{sub-leading}}
\newcommand{\simulat}{\texttt{SIMULATeQCD}}

\begin{document}

%\linenumbers

\title{Curvature of the chiral phase transition line from the magnetic equation of state 
of \boldmath$(2+1)$-flavor QCD}

\author{H.-T. Ding}
\affiliation{ Key Laboratory of Quark \& Lepton Physics (MOE) and Institute of
Particle Physics, Central China Normal University, Wuhan 430079, China}
\author{O. Kaczmarek}
\affiliation{Fakult\"at f\"ur Physik, Universit\"at Bielefeld, D-33615 Bielefeld,
 Germany}
\author{F. Karsch}
\affiliation{Fakult\"at f\"ur Physik, Universit\"at Bielefeld, D-33615 Bielefeld,
 Germany}
\author{P. Petreczky}
\address{Physics Department, Brookhaven National Laboratory, Upton, New York 11973, USA}
\author{Mugdha Sarkar}
\affiliation{Department of Physics, National Taiwan University, Taipei 10617, Taiwan} 
\author{C. Schmidt}
\affiliation{Fakult\"at f\"ur Physik, Universit\"at Bielefeld, D-33615 Bielefeld,
 Germany}
\author{Sipaz Sharma}
\affiliation{Fakult\"at f\"ur Physik, Universit\"at Bielefeld, D-33615 Bielefeld,
 Germany}

\date{\today}

\begin{abstract}

We analyze the dependence of the chiral phase transition temperature on baryon number and 
strangeness chemical potentials
by calculating the leading order
curvature coefficients in the 
light and strange quark flavor basis as well as in the 
conserved charge ($B, S$) basis. Making use of scaling properties
of the magnetic equation of state (MEoS) and
including diagonal as well as off-diagonal contributions in the expansion of the energy-like scaling variable that enters the
parametrization of the MEoS, allows to explore the variation of $T_c(\mu_B,\mu_S)=
     T_c ( 1 - (\kappa_2^B \hmu_B^2
  + \kappa_2^S \hmu_S^2 + 2\kappa_{11}^{BS} \hmu_B \hmu_S))$ along different lines in the $(\mu_B,\mu_S)$ plane. 
On lattices with fixed cut-off in units of temperature, $aT=1/8$, we
find  $\kappa_2^B=0.015(1)$, $\kappa_2^S=0.0124(5)$ and $\kappa_{11}^{BS}=-0.0050(7)$.
We show that the chemical potential dependence along the line of vanishing
strangeness chemical potential
is about 10\% larger than along the
strangeness neutral line. The latter
differs only by about $3\%$ from the
curvature on a line of vanishing 
strange quark chemical potential, $\mu_s=0$.
We also show that close to the chiral limit the strange quark mass contributes like an energy-like variable in scaling relations for pseudo-critical temperatures. The chiral phase transition temperature decreases with
decreasing strange quark mass,
$T_c(m_s)= T_c(m_s^{\rm phy}) (1 -
0.097(2)  (m_s-m_s^{\rm phys})/m_s^{\rm phy}+{\cal O}((\Delta m_s)^2)$.
\end{abstract}

\maketitle

\section{Introduction}\label{sc:intro}
The chiral phase transition in Quantum Chromodynamics (QCD) at finite temperature and vanishing chemical 
potentials is intensively studied in lattice QCD calculations. 
For QCD with physical (degenerate) up
and down quark masses and a strange
quark mass tuned to its physical value,
it is well established that the transition
from the low temperature hadronic phase to a high temperature regime, in which quarks and gluons are the 
dominant degrees of freedom, is continuous and occurs at a pseudo-critical temperature,
$T_{pc}= 156.5(1.5)$~MeV
\cite{Bazavov:2018mes}\footnote{ Pseudo-critical temperatures are not unique. They depend on the observable used to
determine them. The value quoted here 
is obtained by averaging over results obtained with several observables. The spread of results for 
pseudo-critical temperatures is taken into account in the error
quoted. Results obtained from maxima in chiral susceptibilities
only \cite{Bhattacharya:2014ara,Borsanyi:2020fev} are consistent with this result.}. A second order phase transition
occurs in the limit of vanishing up and down 
quark masses, $m_u=m_d=0$, at a temperature $T_c=132^{+3}_{-6}$~MeV
\cite{Ding:2019prx}. A determination of $T_c$ using twisted mass Wilson fermions is consistent with this finding,
$T_c=134^{+6}_{-4}$~MeV \cite{Kotov:2021rah}, while a recent
analytic calculation, using functional renormalization group methods, gives a somewhat larger 
critical temperature, $T_c=142.6$~MeV \cite{Braun:2023qak}.

\begin{figure}
\centering
\includegraphics[width=0.7\columnwidth]{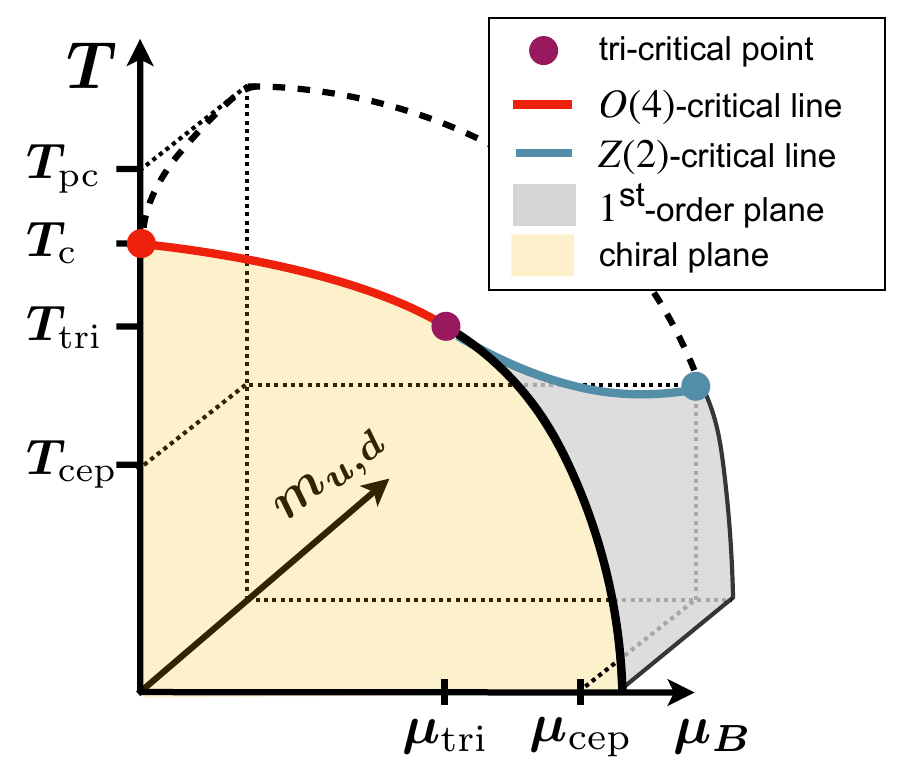}
\caption{Sketch of the QCD phase diagram in the space of temperature
($T$), baryon chemical potential ($\mu_B$) and degenerate light 
quark masses ($m_u=m_d$). See text for a more detailed discussion.}
\label{fig:3dphase}
\end{figure}

In Fig. \ref{fig:3dphase} we show a sketch of the current understanding of the phase diagram in (2+1)-flavor QCD. The existence of a second order phase transition point in QCD at vanishing light quark masses ($m_\ell\equiv m_u= m_d=0)$ and vanishing chemical potentials ($\vec{\mu}=(\mu_u, \mu_d, \mu_s)=0$), or correspondingly at 
vanishing baryon number ($\mu_B$) and strangeness ($\mu_S$) chemical potentials is 
strongly supported by several lattice QCD 
calculations \cite{Ding:2019prx,Cuteri:2021ikv}. However, the existence of a tri-critical point
($T_{tri},\vec{\mu}_{tri})$ in the plane of vanishing light quark masses, $(m_u=m_d=0)$ (yellow plane in Fig.~\ref{fig:3dphase}), is
still based on model calculations \cite{Halasz:1998qr,Hatta:2002sj,Stephanov:2006dn,Buballa:2018hux}. 
At $(T_{tri},\vec{\mu}_{tri})$ the line 
of second order chiral phase transitions emerging from $T_c$, will
meet with a line of first order transitions for $T< T_{tri}$ and 
a line of second order transitions
for $(m_\ell\ne 0,\vec{\mu}\ne 0)$.
A plane of first order transitions (grey plane in Fig.~\ref{fig:3dphase}) is
bounded by these two lines.

The so-called critical endpoint
at $(T_{cep},\vec{\mu}_{cep})$, which is expected to exist at physical values 
of the quark masses, is just one point 
on the line of second order phase transitions emerging from
$(T_{tri},\vec{\mu}_{tri})$. It
belongs to the $3d$, $Z(2)$
universality class.
The nature of the line of second 
order phase transitions, connecting
$(T_c,\vec{\mu}=0)$ and $(T_{tri},\vec{\mu}_{tri})$, is not 
yet settled in detail. Although
the universality class is expected 
to be that of $3d$, $O(4)$ spin models
\cite{Pisarski:1983ms}, which also 
is supported through lattice QCD 
calculations, this is not yet confirmed in detail and a possible 
larger symmetry group cannot be 
ruled out as long as the influence of 
QCD axial symmetry, its 
explicit breaking at vanishing temperature and effective restoration 
in the vicinity of $T_c$, on
the QCD phase transition 
remains controversial \cite{Pisarski:1983ms,Butti:2003nu,Buchoff:2013nra,Pelissetto:2013hqa,Resch:2017vjs,Aoki:2020noz,Ding:2020xlj,Ding:2023oxy,Kovacs:2023vzi}.

Establishing the existence and determining the location of
the critical endpoint at non-vanishing values of quark masses 
and chemical potentials is of great
importance for our understanding of 
the QCD phase diagram and its 
phenomenological implications.
It has been pointed out \cite{Karsch:2019mbv} that the chiral
phase transition temperature, $T_c$, as
well as the tri-critical temperature $T_{tri}$,  
are expected to give upper bounds for the temperature $T_{cep}$ at the possibly existing 
critical endpoint in QCD with physical values of the quark masses and non-vanishing baryon chemical potential. As such it is evident
that it is of interest to also get control over the dependence of 
the chiral phase transition
on chemical potentials \cite{Allton:2002zi}.
It will allow to 
strengthen the bound on $T_{cep}$.

In this work we will analyze the 
influence of non-vanishing chemical potentials $\mu_X$, that couple to the conserved charge operators for net baryon number $(X=B)$ and strangeness $(X=S)$.
We do not include an explicit dependence on an electric charge chemical potential as this
would explicitly violate the
$SU(2)_R\times SU(2)_L$ chiral flavor
symmetry in the light quark sector of QCD.

For non-zero $\mu_B$,
the pseudo-critical as well as the chiral phase
transition temperatures shift to smaller values.
When considering the case of either a single non-vanishing
chemical potential, $\mu_B$, or the case
where the strangeness chemical potential is constrained, e.g. by demanding 
an overall vanishing of the net strangeness density, $n_S=0$,
the shift of pseudo-critical and critical temperatures is, to leading-order, quadratic in $\mu_B$ and is controlled by the so-called curvature coefficient 
$\kappa_2^B$, which has been determined in calculations
with physical (degenerate) light quark masses and a 
physical strange quark mass \cite{Kaczmarek:2011zz,Bonati:2015bha,Bellwied:2015rza,Cea:2015cya,Bazavov:2018mes,Bonati:2018nut,Borsanyi:2020fev}.
In the case of non-zero $\mu_B$ with
$\mu_S,\ \mu_Q$  constrained by demanding 
vanishing strangeness density ($n_S=0$) and a ratio of net electric charge to baryon-number density
that is appropriate for the description of dense media created in heavy ion collisions ($n_Q/n_B=0.4$),
the  curvature coefficient $\kappa_2^B$ is found to be 
$\sim 0.014$ \cite{Bazavov:2018mes,DElia:2018fjp,Borsanyi:2020fev}. Moreover, the next order correction
$\kappa_4^B$ has been found to be consistent with zero
\cite{Bazavov:2018mes,Borsanyi:2020fev}, which
also is in line with recent model calculations
\cite{Biswas:2024xxh}.

When mapping out the critical surface as function of several
non-vanishing chemical potentials, e.g. $T_c(\mu_B,\mu_S)$,
also off-diagonal curvature coefficients, e.g. $\kappa_{11}^{BS}$, need to be taken
into account. We will discuss here 
the dependence of various curvature coefficients on the value of two
degenerate light quark masses with the 
strange quark mass tuned to its 
physical value, aiming at an extrapolation to the chiral limit. 
In this study we do not yet try to
extrapolate to the continuum limit. All calculations are performed at finite values of the lattice
cut-off. We will determine diagonal and off-diagonal curvature coefficients,
$\kappa_2^B$, $\kappa_2^S$ and $\kappa_{11}^{BS}$,
in the conserved charge basis, $(\mu_B,\mu_S)$, as well as the related curvature coefficients $\kappa_2^\ell$, $\kappa_2^s$ and $\kappa_{11}^{\ell s}$, in the
quark flavor basis $(\mu_\ell,\mu_s)$. 

Close to the chiral limit, universal scaling relations allow the determination of the chiral phase transition
temperature and its dependence on, e.g., the chemical potentials,
which are a particular set of energy-like scaling fields
coupled to energy-like observables \cite{Clarke:2020htu}. 
We will use a scaling ansatz for chiral observables to
extract the curvature coefficients $\kappa_2^X$ and $\kappa_{11}^{XY}$, ignoring 
regular or sub-leading universal, correction-to-scaling contributions. This ansatz becomes exact when approaching
the chiral limit. 

This paper is organized as follows.  
In Section II we present some details
on the calculational set up used for this
work. In Section III we introduce several chiral observables and discuss universal properties of (2+1)-flavor QCD near the
chiral limit. Section IV gives an update 
on the determination of the chiral 
phase transition temperature at vanishing chemical potentials, using three sets of susceptibilities that define three different sets of pseudo-critical temperatures. In this section we also present a determination of the non-universal scale parameter $z_0$, which is needed to define the leading universal part of chiral observables entering the magnetic equation of state. 
In Section V we present our results on curvature coefficients calculated in conserved charge 
and conserved flavor basis. Section VI is devoted to a discussion of the
dependence of the chiral phase transition temperature
on the value of the strange quark mass. 
We finally give our conclusions in Section VII. In three appendices we
give some details on finite volume effects
in the vicinity of the chiral phase transition 
temperature (App.~\ref{app:finitevol}),
a discussion about
the sub-leading correction-to-scaling
terms that contribute to all thermodynamic
observables close to the chiral limit (App.~\ref{app:cts})
and present the determination
of the leading order coefficient $s_1(T,H)$ that relates the strangeness and baryon chemical potentials in strangeness neutral systems (App.~\ref{app:s1}).

\section{Calculational setup}
 
We consider (2+1)-flavor QCD 
with a strange quark mass $m_s$ tuned to its physical value and degenerate
up and down quark masses, ($m_\ell\equiv m_u= m_d$), that 
 shall be changed to smaller-than-physical values, 
 starting from $m_\ell/m_s=1/20$, including the physical 
 value, $m_\ell/m_s=1/27$ and reaching the smallest quark mass ratio $m_\ell/m_s=1/160$. 
 We introduce the
 chemical potentials $\mu_f$ for each quark flavor.
 The partition function, defined
 on a $4d$ Euclidean space-time lattice of size $N_\sigma^3N_\tau$, is given as
  \begin{equation}
   Z(T,\vec{\mu},V) =\hspace*{-0.1cm} \int \mathcal{D}U e^{-S_G(\beta,U)}\hspace*{-0.2cm} \prod_{f=u,d,s}(\det \Mc_f(m_f,\mu_f))^{1/4} ~,
  \end{equation}  
where $\vec{\mu}=(\mu_u,\mu_d,\mu_s)$ and $\Mc_f$ denotes the staggered fermion matrix for quark flavor $f$;
$Z(T,\vec{\mu},V)$, of course, also is a function of the quark masses $m_f$, 
which we have not specified explicitly.
Temperature $T$ and volume $V$ are 
related to the temporal ($N_\tau\equiv (Ta)^{-1}$)
and spatial ($N_\sigma\equiv V^{1/3}/a$) extent of the lattice
with lattice spacing $a(\beta)$ 
being controlled by the gauge
coupling\footnote{The gauge coupling $\beta$, introduced here, should not be confused with the critical exponent $\beta$
that is used frequently in this work.} $\beta$.

\begin{table*}[t]
\begin{center}
\begin{tabular}{|ccc|rrrr|}
\hline
~&~&~&~~$H=1/27$&~~$H=1/40$&~~$H=1/80$&~~$H=1/160$
\\
~&~&~&$~~N_\sigma = 32$ &$~~N_\sigma = 40$
&$~~N_\sigma = 56$
&$~~N_\sigma = 56$
\\
~&~&~&$~~z_{L,b} = 0.94$ &$~~z_{L,b} = 0.88$
&$~~z_{L,b} = 0.84$
&$~~z_{L,b} = 1.10$
\\
\hline
$\beta$ & $\tilde{m}_s$   & ~$T$[MeV]~ & \multicolumn{4}{c|}{\#configurations} \\ \hline
\hline
6.245 & 0.0830 & 134.84 & 123004 & --& -- &-- ~~\\
6.260 & 0.0810 & 136.98 & -- & 71824 & -- &-- ~~\\
6.285 & 0.0790 & 140.62 & 145814 & 71547 & 28703 & 13254 ~~\\
6.300 & 0.0772 & 142.85 & -- & 71514 & 28464 & 13299 ~~\\
6.315 & 0.0759 & 145.11 & 145179 & 71571 & 31754 & 15693 ~~\\
6.330 & 0.0746 & 147.40 & -- & 70212 & 31751 & 15634 ~~\\
6.354 & 0.0728 & 151.14 & 125679 & 51135 & 31561 & 13200 ~~\\
6.365 & 0.0716 & 152.88 & -- & 49879 & -- &-- ~~\\
6.372 & 0.0711 & 154.00 & -- & --  & 17222 & 13790 ~~\\
6.390 & 0.0694 & 156.92 & 97526 & 51707 & 17794 & 10129 ~~\\
6.423 & 0.0670 & 162.39 & 131200 & 52051 & 10917 & 9137 ~~\\
6.445 & 0.0652 & 166.14 & 176361 & 50689 & 10863 & 9479 ~~\\
6.474 & 0.0632 & 171.19 & 232116 & 28080 & -- &-- ~~\\
6.500 & 0.0614 & 175.84 & 151606 & 29505 & -- &-- ~~\\
\hline
\end{tabular}
\end{center}
\caption{Run parameter for simulations on lattices with temporal extent $N_\tau=8$ and spatial lattice size $N_\sigma^3$. We used 
$\tilde{m}_s=m_sa$ for the dimensionless strange quark mass times lattice spacing {\it a}.
The stored gauge field configurations for $H=1/27$ are separated by 10 time units in rational hybrid Monte Carlo calculation while for smaller $H$, they are separated by 5 time units. The finite-size scaling
variable $z_{L,b}$ is defined in Eq.~\ref{zLb}.}
\label{table_allml}
\end{table*}

For the purpose of the calculations presented here we substantially extended the set of gauge field ensembles generated previously at smaller-than-physical light quark masses for the determination of the chiral phase transition temperature \cite{Ding:2019prx}.
We use a temperature independent normalization for the chiral
symmetry breaking parameter,
which is proportional to the degenerate light quark masses
$m_\ell$, {\it i.e.} we use 
the light quark mass in units
of the strange quark mass, $H=m_\ell/m_s$.
Some preliminary results on the curvature of the chiral phase transition line in the limit $H\rightarrow 0$ have been presented already in \cite{Kaczmarek:2021ufg}.

All new calculations have
been performed on lattices with temporal extent
$N_\tau=8$ and several spatial lattice sizes with aspect ratio
$N_\sigma/N_\tau$ varying between 4 and 7. We vary $N_\sigma$ in order to ensure 
that the unnormalized (bare) finite-size scaling variable, 
\begin{equation}
z_{L,b}=\frac{N_\tau}{N_\sigma H^{\nu/\beta\delta}} \; ,
\label{zLb}
\end{equation}
stays small. In our
calculations we have $z_{L,b} < 1$,
except for the smallest quark mass 
ratio $H=1/160$, where $z_{L,b}=1.1$.
In our previous analysis
of critical behavior in (2+1)-flavor
QCD \cite{Ding:2019prx} the region
$z_{L,b} < 1$ has been found to
correspond to a region where finite
volume effects are small. 
Further discussion of finite volume effects in (2+1)-flavor QCD, that gives support to this statement, is given in Appendix \ref{app:finitevol}. 
We discuss the possible influence
of finite volume effects in our 
data for $H=1/160$ in the next sections.
In terms of the pion mass these values for
$z_{L,b}$ correspond to physical volumes of 
size $m_\pi V^{1/3}\simeq 3-4$.
The values for $z_{L,b}$ are also
given in Table~\ref{table_allml}.

The gauge field ensembles, used in our study, have
been generated with the
Highly Improved Staggered Quark (HISQ) action
\cite{Follana:2006rc} with tree-level
coefficients
and a tree-level Symanzik improved gauge action,
previously used by us also in calculations with a physical strange
quark mass and two degenerate,
physical light quark masses \cite{Bazavov:2011nk,Bollweg:2021vqf}.
The ensembles were generated with a strange quark mass tuned to its physical value and five 
values of the degenerate light 
quark masses corresponding to light 
to strange quark mass ratios,
$H=1/20$, 1/27, 1/40, 1/80 and 1/160. 
In the continuum limit these quark mass ratios 
correspond to light pseudo-scalar pion masses,
$m_\pi \simeq 160$~MeV,\ 140 MeV, 110 MeV, 80 MeV and 55 MeV, respectively. 
Using this set-up on lattices with temporal extent $N_\tau=8$, the chiral phase transition temperature has been
determined previously to be $T_c^{N_\tau=8} = 144(2)$~MeV \cite{Ding:2019prx}. We give an 
update on this value in Sec.~\ref{sc:pseudo}.

For scale setting we use a parametrization of the $\beta$-function of (2+1)-flavor QCD with physical light and strange quark masses given in \cite{Bollweg:2021vqf}. The scale is set using
the kaon decay constant $f_K=155.7/\sqrt{2}$~MeV \cite{FlavourLatticeAveragingGroup:2019iem}.

We summarize details on our 
simulation parameters and the number of 
gauge field configurations used in these calculations in 
Table~\ref{table_allml}. Data for 
the largest quark mass ratio, $H=1/20$, have been taken from
Table XI of \cite{Bazavov:2011nk}.

All our numerical calculations are 
done at vanishing values of the chemical potentials using the simulation package $\simulat$~\cite{Bollweg:2021cvl,HotQCD:2023ghu}.
Thermodynamic observables at non-vanishing values of the 
chemical potentials are calculated using the Taylor expansion approach
\cite{Gavai:2003mf,Allton:2005gk,Bazavov:2018mes}.

\section{Chiral observables}

\label{sc:observables}

\subsection{Chiral order parameter}
Starting point for our determination of the chiral phase transition temperature and its dependence on quark masses and various chemical
potentials is a set of unrenormalized chiral condensates ($\tilde{\Sigma}_f$) for flavor $f=u, d,s$, 
and the related  chiral susceptibilities ($\tilde{\chi}_{m,f}$),
\begin{eqnarray}
    \tilde{\Sigma}_f&=& \frac{T}{V}\frac{\partial \ln Z}{\partial m_f} \; ,
    \label{Sigma}\\
     \tilde{\chi}_{m,f} &=& \frac{\partial \tilde{\Sigma}_f}{\partial m_f} 
\label{chi-l}\; . 
\end{eqnarray}
We also introduce the unrenormalized 2-flavor light quark condensate and its susceptibility as the sum of the corresponding up and down quark chiral observables, {\it i.e.}
  \begin{eqnarray}
  \tilde{\Sigma}_\ell &=& \tilde{\Sigma}_u
  +\tilde{\Sigma}_d \;\; ,
  \label{Sigmaell} \\
  \tilde{\chi}_{m,\ell} &=& \left(
  \frac{\partial}{\partial m_u} + \frac{\partial}{\partial m_d}\right)
  \tilde{\Sigma}_\ell \;\; .
  \label{chiell}
  \end{eqnarray}

The chiral condensates require multiplicative and additive renormalization to be well defined
in the continuum limit. We take care
of the multiplicative renormalization
by multiplying the light quark chiral 
condensate with the strange quark 
mass. Furthermore, we divide by 
appropriate powers of $f_K$ to define
a dimensionless order parameter and
its susceptibility,
\begin{eqnarray}
M_\ell &=& \frac{m_s}{f_K^4}  \tilde{\Sigma}_\ell\; , \label{MfK}\\
\chi_{\ell} &=& \frac{m^2_s}{f_K^4}  \tilde{\chi}_{m,\ell}\; .
\label{chifK}
\end{eqnarray}
Additive UV divergences in the chiral condensates are proportional to the quark masses. They can be eliminated
by subtracting a suitable observable that
contains the same UV divergence. Commonly
used is the so-called 
2-flavor subtracted chiral condensate
($M_{\rm sub}$)
\cite{Cheng:2007jq,Ejiri:2009ac}, which is 
obtained by subtracting an appropriate fraction of the strange quark condensate from the light quark condensate,
    \begin{equation}
      M_{\rm sub} = M_\ell - 2 H \frac{m_s}{f_K^4}
      \tilde{\Sigma}_s\; .
\label{ordersub}
  \end{equation}
Another possibility to introduce an order parameter for chiral symmetry restoration
is to subtract from the light quark condensate a fraction of
the corresponding light quark susceptibility \cite{Unger:2010wcq,Kotov:2021rah},
\begin{equation}
M\equiv M_\ell -H \chi_\ell \; .
\label{order}
\end{equation}
This version of a renormalized order
parameter has the advantage of
not involving explicit contributions from
the strange quark condensate, which
potentially contains larger regular
contributions. Moreover, $M$ is directly related to a combination of universal
scaling functions, which leads to pseudo-critical temperatures
that have a weaker quark mass dependence than the chiral
susceptibility or other mixed susceptibilities. 
In the following we will use the
unrenormalized order parameter $M_\ell$
as well as the renormalized order parameters $M_{\rm sub}$ and $M$ to
define various renormalized susceptibilities that can be used to 
define pseudo-critical temperatures at
non-vanishing $H$.

Finally, we also introduce a mixed susceptibility that is
sensitive to changes in the light quark condensate arising
from variations of the strange quark mass,
\begin{eqnarray}
\chi_{\ell,m_s} &=& \frac{m^2_s}{f_K^4} 
\left. \frac{\partial \tilde{\Sigma}_\ell}{\partial m_s}\right|_{m_s^{\rm phy}} 
 \; .
\label{chils}
\end{eqnarray}

\subsection{Chiral and mixed susceptibilities}

For the determination of the dependence
of the chiral phase transition temperature, $T_c$, on chemical
potentials we will calculate pseudo-critical temperatures that are
defined as the location of maxima in various susceptibilities.

As the quadratic UV divergence, present in
the unsubtracted order parameter $M_\ell$,
neither depends on temperature nor on
chemical potentials, these divergences
are removed when considering susceptibilities obtained
as derivatives with respect to $T$
or chemical potentials, respectively.

We thus consider in the following 
several mixed susceptibilities obtained from 
derivatives of $M_\ell$ with respect to $T$ or $\hmu_f=\mu_f/T$ with $f=\ell$ or $s$,
\begin{eqnarray}
 \chi^{M_\ell}_{t(T)}
&=&  -T_c \frac{\partial M_\ell}{\partial T}
\; , 
\label{chi-t}  \\
 \chi^{M_\ell}_{t(f,f)}
&=& - \frac{\partial^2 M_\ell}{\partial \hmu_f^2} 
\;\; ,
\label{chi-f} \\
 \chi^{M_\ell}_{t(\ell, s)}
&=& - \frac{\partial^2 M_\ell}{\partial \hmu_\ell \partial \hmu_s}
\; ,
\label{chi-mu}
\end{eqnarray}
Here $t(T)$ and $t(f,g)$
indicate that derivatives of the order parameter with respect to $T$ or temperature like 
variables $\hmu_f$ and $\hmu_g$
are taken.
A corresponding set of susceptibilities 
can be defined by replacing $M_\ell$
by the renormalized order parameter $M_{\rm sub}$ or $M$. We will use in
the following the mixed
susceptibility
\begin{eqnarray}
\chi^M_{t(T)}
&=&  -T_c \frac{\partial M}{\partial T}
\; . \label{chi-tM} 
\end{eqnarray}

Related to the two versions of a renormalized 
chiral order parameter, $M_{\rm sub}$
and $M$, we
introduce two versions of 
chiral susceptibilities,
{\it i.e.} the
derivatives of $M_{\rm sub}$ or
$M$ with 
respect to the light quark mass $m_\ell$,
\begin{eqnarray}
\chi_{m}^{M_{\rm sub}}
&=& \left(
  \frac{\partial}{\partial m_u} + \frac{\partial}{\partial m_d}\right) M_{\rm sub}
\; , \label{chisub} \\
\chi_{m}^{M}
&=& \left(
  \frac{\partial}{\partial m_u} + \frac{\partial}{\partial m_d}\right) M
\; .  \label{chiM}
\end{eqnarray}
In the following we will only make use of
$\chi_{m}^{M_{sub}}$ as the calculation
of $\chi_{m}^{M}$ would require the
calculation of three derivatives of the 
$\ln Z$ with respect to the light quark masses.

\subsection{Universal critical behavior}  
\label{sc:universal} 
At vanishing values of the 
chemical potentials the existence of a 
continuous $2^{nd}$
order phase transition, occurring at vanishing values of the two degenerate light quark masses, $m_\ell$, has been 
established \cite{Ding:2019prx,Cuteri:2021ikv}. As the
chemical potentials $(\mu_\ell,\mu_s)$
do not explicitly break the chiral symmetry, 
the point $(\mu_\ell, \mu_s)=(0,0)$
is part of a surface of $2^{nd}$
order phase transitions that occur at temperatures, $T_c(\mu_\ell,\mu_s)$.
In the vicinity of this critical surface the free energy density, 
$f= - (T/V)\ln Z$, can be split into a
singular contribution and  sub-leading corrections that are of relevance in some range of $H\ne 0$,
  \begin{equation}
 f = f_s(T,\vec{\mu},\vec{m}) + f_{\rm sub-lead}(T,\vec{\mu},\vec{m}) \; ,  
 \label{free}
 \end{equation}
with $\vec{m}=(m_\ell,m_s)$ and $\vec{\mu}=(\mu_\ell,\mu_s)$. The singular part gives rise to divergences in higher order derivatives of the free energy density. The 
sub-leading corrections involve non-singular, regular terms as well 
as sub-dominant, universal singular corrections-to-scaling (cts).

In the vicinity of the critical surface the leading singular contribution to the free energy density dominates the behavior of the chiral and mixed susceptibilities.
The dominant singular contribution
is written in terms of energy-like and magnetization-like scaling fields, $u_t$ and $u_h$, respectively. The former couples
to operators in the QCD Lagrangian, which are
invariant under chiral transformations in the 
light, degenerate 2-flavor sector and is 
a function of all combinations of couplings (parameters) appearing in the QCD Lagrangian, which leave the Lagrangian invariant under chiral rotations. 
The latter, on the other hand, depends
on combinations of couplings that
break this symmetry.
The scaling fields $u_t$ and $u_h$
vanish at a critical point. In its
vicinity they may be expanded
in a Taylor series. Usually one
uses only the leading order Taylor
series expansion, $u_t=t+{\cal O}(t^2,t h^2)$, $u_h=h+{\cal O}(t h)$, with
  \begin{eqnarray}
  t &=& \frac{\bar{t}}{t_0}= \frac{1}{t_0} \left( \Delta T + \kappa_2^\ell \hmu_\ell^2 
  + \kappa_2^s \hmu_s^2 + 2\kappa_{11}^{\ell s} \hmu_\ell \hmu_s
  \right), \label{coupling-t} \\
  h &\equiv& \frac{H}{h_0} = \frac{1}{h_0} \frac{m_\ell}{m_s}. 
  \label{coupling-h}
  \end{eqnarray}
Here $t_0$, $h_0$ are dimensionless non-universal constants just like $T_c$,
and
\begin{equation}
    \Delta T = \frac{T-T_c}{T_c} \; .
\end{equation}

In Eq.~\ref{coupling-t} we introduced
the reduced temperature $t$ as function
of the energy-like couplings $T$ and
chemical potentials in the flavor basis.
This may as well be done
in the conserved charge basis using
the chemical potentials $(\mu_B, \mu_S)$.
Note that we do not include the isospin or electric charge chemical potential in $\bar{t}$ since 
this amounts to introducing independent up and down quark chemical potentials which
explicitly breaks the symmetry group to 
$U(1)_u \times U(1)_d \times U(1)_s$.

Up to the sub-leading contributions from corrections-to-scaling and contributions from regular terms,
the temperature and mass dependence of the singular part of the free energy density
is controlled by a single scaling variable $z$, 
\begin{equation}
  f_s(T,\vec{\mu},\vec{m})=h_0 h^{1+1/\delta} f_f(z)\; ,
  \label{fs}
\end{equation}
with $z= z_0 z_b$ and
\begin{eqnarray}
  z_b=\bar{t}/H^{1/\beta\delta}\; &,& \; z_0 = h_0^{1/\beta\delta}/t_0 \; .
  \label{zb}
 \end{eqnarray}
The critical exponents $\beta$, $\delta$ are unique in the
universality class of the phase transition.
As we consider in the following 
the chiral limit taken at fixed lattice cut-off, we use for definiteness critical
exponents of the $3d$, $O(2)$ universality class, {\it i.e.} we use
\cite{Engels:2000xw},
  \begin{equation}
      \beta=0.3490(30)\; ,\; \delta=4.7798(5)
      \; , \;
      \omega\nu_c =0.32(1)
      \; .
  \end{equation}
Here we give in addition to the 
two critical exponents $\beta, \delta$ also the 
combination $\omega\nu_c\equiv
\omega \nu/\beta\delta$, with
$\nu=\beta(1+\delta)/3$ and
$\omega$ denoting the sub-leading universal correction-to-scaling exponent
\cite{Hasenbusch:1999cc,Engels:2000xw}.
The combination $\omega\nu_c$ controls the cts contribution to the order parameters $M$ or $M_\ell$, respectively. 

The scaling functions $f_G(z)$ and
$f_\chi(z)$, which control the universal, singular parts of the order parameter and its susceptibilities, are related to $f_f(z)$,
  \begin{eqnarray}
      f_G(z) &=& -\left(1+\frac{1}{\delta}\right)f_f(z) + \frac{z}{\beta\delta}f_f^\prime (z) \; ,
      \label{fG} \\
      f_\chi(z) &=& \frac{1}{\delta}\left( f_G(z) - \frac{z}{\beta} f'_G(z)\right)\; .
      \label{fchi}
  \end{eqnarray}
Taking into
account the temperature independent normalization, which we introduced 
in Eqs.~\ref{MfK}, \ref{chifK}, we
get for the unsubtracted order parameter, $M_\ell$, as well as the renormalized
order parameter $M_{\rm sub}$ introduced in Eq.~\ref{ordersub},
\begin{eqnarray}
M_{\ell /{\rm sub}} &=&  h^{1/\delta} f_G(z) + ~\subl
      \; . \label{M-scaling} 
      \end{eqnarray}
The universal part of the order parameter $M$, introduced in
Eq.~\ref{order}, is related to
the difference of the scaling functions $f_G(z)$ and $f_\chi(z)$,
\begin{equation}
    M = h^{1/\delta} \left( f_G(z)-f_\chi(z) \right) + ~\subl \; .
\end{equation}
Correspondingly the mixed susceptibilities, obtained 
as temperature derivatives of these
order parameters, have different 
universal, singular parts,
      \begin{eqnarray}
      \chi^{M_\ell}_{X} &=& -a_X t_0^{-1} h^{(\beta-1)/\beta\delta} f'_G(z)  + ~\subl \; ,\; \nonumber \\
      && \hspace{2.7cm}X=t(T),t(f,f),t(\ell, s)\;,
      \label{chitm-scaling}
      \\
    \chi^M_{t(T)} &=& -t_0^{-1} h^{(\beta-1)/\beta\delta}
     (f'_G(z)-f'_\chi(z)) + ~\subl \; . \nonumber
     \label{chitM-scaling}
  \end{eqnarray}
Here $a_X= 1, 2\kappa_2^f, 2\kappa_{11}^{\ell  s}$, respectively.
Furthermore, the singular part of the chiral susceptibility $\chi^{M_{\rm sub}}_m$, 
introduced in Eq.~\ref{chisub},
is controlled by the scaling function $f_\chi(z)$,
 \begin{equation}
     \chi^{M_{\rm sub}}_m =  \frac{h^{1/\delta-1}}{h_0} f_\chi(z) + ~\subl  \; .
      \label{chil-scaling}
 \end{equation}

Close to $T_c$ contributions arising from corrections-to-scaling 
give the dominant sub-leading contribution to the order parameter.
This contribution is proportional to $H^{1/\delta+\omega\nu_c}=H^{0.53(1)}$,
while regular terms are
proportional to $H$. The correction-to-scaling term thus
leads to a sub-leading, still divergent contribution, in susceptibilities.

Close to the chiral limit, where corrections-to-scaling and regular
 contributions can be neglected, the
 pseudo-critical temperature, deduced
 from the maximum of $\chi^{M_{\rm sub}}_m(T,H)$, 
 is related to the maximum of $f_\chi(z)$. The 
 maxima of the mixed susceptibilities
 obtained from derivatives of $M_\ell$
 or $M$ with respect to temperature-like
 couplings, on the other hand, 
 are related to maxima of the scaling functions $-f'_G(z)$ and $f'_\chi(z)-f'_G(z)$, respectively.
 
 These maxima occur at
 specific values of the scaling variable $z$, which we denoted by $z_m$ for the maximum
 of $f_\chi(z)$,  $z_t$ for that of $-f'_G(z)$, and $z_{t,M}$ for the 
 maximum of the difference $f'_\chi(z)-f'_G(z)$.
 Results for these $z$-values in the
 $3d$, $O(2)$ universality class 
 are obtained from the scaling functions
 given in
 \cite{Karsch:2023pga},
 \begin{eqnarray}
 z_m &=& 1.6675(68) \; , \label{zzp}\\  
 z_{t} &=& 0.7991(96) \; , \label{zzt} \\
 z_{t,M} &=& 0.629(10) \; .
 \label{zztM}
  \end{eqnarray}

These specific values of $z$ then define
pseudo-critical temperatures, obtained from the maxima of the magnetization-like susceptibility $\chi^{M_{\rm sub}}_m$ or those obtained
from the mixed susceptibilities ($\chi^{M_\ell}_t(T),\ 
\chi^{M_\ell}_{t(\ell,\ell)},
\chi^{M_\ell}_{t(\ell,s)})$,
as well as $\chi^M_{t(T)}$. They give
rise to three sets of pseudo-critical
temperatures. Their leading, universal
parts are given by 
 \begin{eqnarray}
 T_{pc,x}(\hmu_\ell,\hmu_s,H) &=& T_c 
\Bigl( 1 + 
(\kappa_2^\ell \hmu_\ell^2
  + \kappa_2^s \hmu_s^2 + 2\kappa_{11}^{\ell s} \hmu_\ell \hmu_s)
   \nonumber \\
&&+ \frac{z_{x}}{z_0}H^{1/\beta\delta} \Bigr)
\label{Tpcx} \; ,\; x=m, t, (t,M)\; .
 \end{eqnarray}

 For $H=0$ Eq.~\ref{Tpcx} defines the surface
 of critical temperatures, $T_c(\hmu_\ell, \hmu_s)\equiv T_{pc,x}(\hmu_\ell,\hmu_s,0)$, spanned by 
 the light and strange quark chemical potentials, respectively. 

\begin{figure*}[t]
\includegraphics[width=0.47\textwidth]{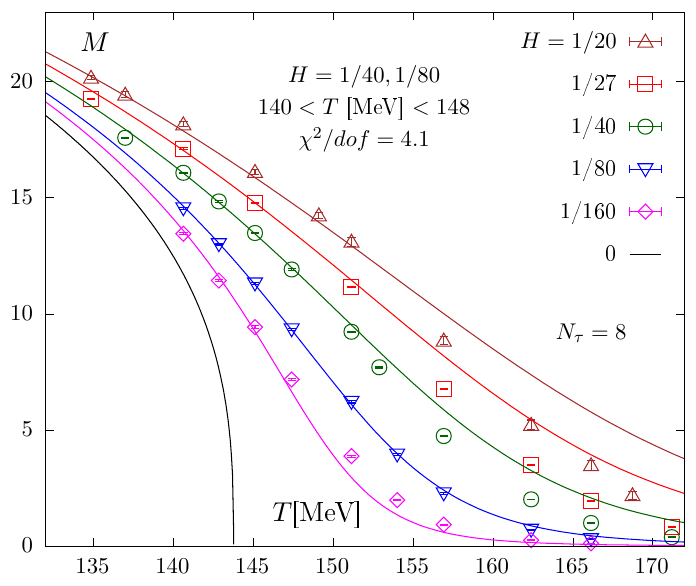}
\includegraphics[width=0.49\textwidth]{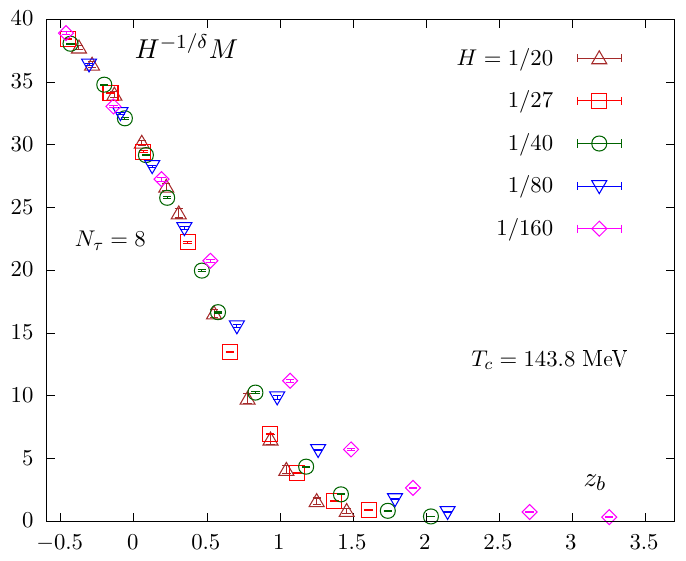}
\caption{{\it Left:} The dimensionless renormalized order parameter $M$ versus
$T$. Shown also is 
a fit to the data for $H=1/40$ and $1/80$ in
the temperature interval $T\in [140~{\rm MeV}:148~{\rm MeV}]$. The resulting fit
parameters
$(T_c, z_0, h_0^{-1/\delta})=(143.8(2){\rm MeV},1.45(3),39.0(3))$,
are also
given in Table~\ref{tab:Tpc-fits}.
The fit result is also shown beyond the actual fit range.
{\it Right:}
The renormalized order parameter $M$
versus the bare scaling variable $z_b$ calculated
using as input only the critical temperature from the fit shown
in the left hand figure.
}
\label{fig:Sigma}
\end{figure*}

 \section{Magnetic equation of state and pseudo-critical temperatures at \boldmath$\vec{\mu}=0$}
\label{sc:magnetic}

As we will exploit the structure 
of the singular part of the QCD
free energy to determine the
dependence of the chiral phase transition temperature
on chemical potentials, we need as 
input the non-universal scales $z_0$ and $T_c$ at $(\mu_\ell,\mu_s)=(0,0)$.
These have been determined previously \cite{Ding:2019prx}. 
We will improve here our analysis on
lattices with temporal extent $N_\tau=8$ making use of our increased
statistics at several parameter values.

We will analyze the scaling 
behavior of the
renormalized  order parameter $M$ and extract pseudo-critical temperatures from the chiral and various mixed
susceptibilities introduced in the previous sections.

 \subsection{Renormalized order parameter}

In Fig.~\ref{fig:Sigma}~(left) we show results for the renormalized order parameter $M$ introduced in Eq.~\ref{order}.
Results have been obtained at
five different values of $H$ on lattices
with temporal extent $N_\tau=8$. We fitted these data 
to the universal scaling ansatz
\begin{equation}
M = h_0^{-1/\delta} H^{1/\delta} (f_G(z)-f_\chi(z))\; ,
\label{fitansatz}
\end{equation}
with $f_G(z)$ and $f_\chi(z)$ 
denoting scaling functions in the $3d$, $O(2)$ universality class \cite{Karsch:2023pga}. The scaling functions used in the fit are expressed in the Schofield parametrisation obtained in \cite{Karsch:2023pga}.
This ansatz involves three fit parameters, ($T_c,z_0,h_0^{-1/\delta}$).
As we do not include contributions
from regular or sub-leading universal terms in this fit we need to
restrict the fit to small
quark masses and a 
temperature region close to the pseudo-critical
temperature. This has also been done in earlier analyses of the 
magnetic equation of state \cite{Ejiri:2009ac}. It should be noted that in regular contributions to $M$ the leading $H$ dependent term gets cancelled, leaving only a weaker $H^3$ dependent contribution arising from regular terms.

We fitted the scaling 
ansatz, Eq.~\ref{fitansatz}, to data for $M$ obtained 
with light to strange quark mass ratios $H=1/40$ and $1/80$ in the 
temperature interval $T\in [140~{\rm MeV}:T_{\rm max}]$ with $T_{\rm max}=146~{\rm MeV}$ and $148~{\rm MeV}$, respectively. 
These fits have been performed with 
and without including the data for
the smallest quark mass $H=1/160$, which
have been obtained on our smallest physical volume and may still suffer somewhat from finite volume effects.
The resulting fit parameters are
given in Table~\ref{tab:Tpc-fits}.
As can be seen the fit parameters vary little, although the $\chi^2/dof$
of the fits is quite sensitive to
the chosen fit-interval and the range
of $H$-values used in the fit.

In Fig.~\ref{fig:Sigma}~(right) we show the rescaled order 
parameter $M$ as function of the scaling variable $z_b$ introduced 
in Eq.~\ref{zb}. As can be seen, scaling holds well at 
least up to $z_b\simeq 0.5$. For
our smallest quark mass ratio, $H=1/160$, this corresponds to 
a temperature interval
$(T-T_c)/T_c\simeq 0.026$, which is
similar to that finally used also
in \cite{Ejiri:2009ac}.

\begin{figure}[t]
\includegraphics[width=0.48\textwidth]{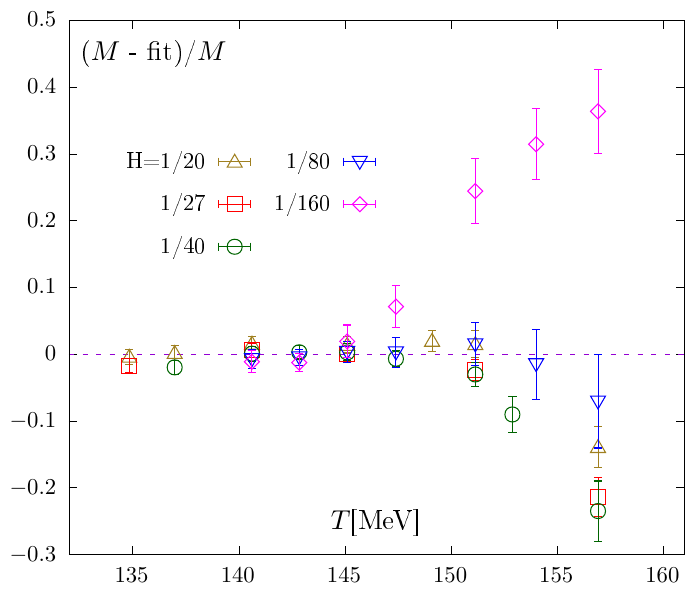}
\caption{Relative deviation of data from
the fit shown in Fig.~\ref{fig:Sigma}~(left). Deviations are shown also outside the actual fit range and for values of $H$ not included in the fit.}
\label{fig:Sigma-deviations}
\end{figure}

\begin{table}[htb]
\begin{center}
\begin{tabular}{|c|c||c|c|c|c|c|}
\hline
$H^{-1}$ & $T~[{\rm MeV}]$& $T_c~{\rm [MeV]}$ & $z_0$ &$h_0^{-1/\delta}$
&$\chi^2/dof$\\
\hline
(40,80) & [140:146]&143.6(1) &1.39(2) & 39.4(2) & 1.2\\
(40,80) &[140:148]&143.8(2)& 1.45(3)& 39.0(3)& 4.1 \\
(40,80,160)&[140:146] & 143.6(1)& 1.36(3)&
39.4(2) & 3.7 \\
(40,80,160)&[140:148] & 143.8(2)& 1.40(4)&
38.8(4) & 15.1 \\
\hline
\end{tabular}
\end{center}
\caption{Fit parameter obtained for fits with different sets of quark masses and in different temperature intervals. The fit shown in
Fig.~\ref{fig:Sigma}~(left) corresponds to the parameter set with a $\chi^2/dof=4.1$.}
\label{tab:Tpc-fits}
\end{table}

The fits performed in a small temperature interval and for small values of $H$ 
still provide a good description of our data sets for
larger and smaller masses as well as for
data outside the temperature range used in the fits. 
Deviations of data outside the fit range from the fit prediction provide an estimate 
for the influence of regular 
or sub-leading universal 
contributions. 
In Fig.~\ref{fig:Sigma-deviations} we show the relative deviation of 
data from the fit also outside the actual fit interval. This suggests
that corrections to universal scaling behavior arising from regular 
or sub-leading universal terms remain smaller than 10\%  for 
$(T-T_c)/T_c\lsim 0.06$.

In the following we use the average of 
the fit results for $(T_c, z_0,h_0^{-1/\delta})$ obtained
by leaving out the data for $H=1/160$
in the fit. We take care of this data
set by including the differences as systematic error contributing to the
quoted error for $(T_c, z_0,h_0^{-1/\delta})$. We use,
\begin{eqnarray}
    T_c &=& 143.7(2)~{\rm MeV}\; ,  \label{Tcfix}\\
    z_0 &=& 1.42(6) \; , \label{z0fix} \\
    h_0^{-1/\delta} &=& 39.2(4)\; . \label{h0fix}
\end{eqnarray}

Using the fit result for the scale parameter $z_0$, 
we also conclude from the rescaled order parameter data, 
shown in Fig.~\ref{fig:Sigma}~(left),
that the parameter
range in which we find good scaling behavior without
including sub-leading corrections in our fits,
corresponds to the region
$|z|\lsim 0.7$. This suggests
that the peak positions of 
mixed susceptibilities are only
mildly influenced by contributions from sub-leading corrections to the dominant universal scaling behavior, as $z_t$, $z_{t,(t,M)}$ are of similar magnitude (see Eqs.~\ref{zzt}, \ref{zztM}). On the other hand, their influence on the location of the peak of the chiral susceptibility will be larger, as the peak is located at $z_m\simeq 1.7$ (Eq.~\ref{zzp}). We will
analyze this in more detail in the next subsection.

\begin{table*}[htb]
\begin{center}
\begin{tabular}{|cc|c|c|c|c|c|}
\hline
~&~ &\multicolumn{5}{|c|}{Pseudo-critical temperatures in units of MeV} \\ 
\hline
\multicolumn{2}{|c|}{observable}&$H=1/20$&$H=1/27$&$H=1/40$&$H=1/80$&$H=1/160$ \\[2pt]
\hline
(i)&$~~\chi^{M_{\rm sub}}_m~$~~& ~164.63(5)~ & ~160.99(12)~&~157.89(27)~&~153.91(30)~&~150.93(34)~ \\[2pt]
\hline
(ii)&$\chi^{M_\ell}_{t(T)}$& 162.84(28) & 159.76(12)&155.66(09)&151.29(14)&147.92(14) \\[2pt]
~&$~~\chi_{\ell,m_s}$~~& -- & ~159.39(22)~&~156.38(25)~&~150.51(63)~&~148.48(74)~\\[2pt]
~&$\chi^{M_\ell}_{t(\ell,\ell)}$&-- & 159.94(32) & 156.07(33) & 150.7(1.6) & 148.7(1.2) \\[2pt]
~&$\chi^{M_\ell}_{t(s,s)}$&-- & 162.76(26) & 159.32(29) & 153.6(96) & 149.87(54) \\[2pt]
\hline
(iii)&$~~\chi^M_{t(T)}$~~& ~156.40(68)~ & ~153.44(11)~&~150.70(06)~&~148.04(12)~&~145.85(17)~\\[2pt]
\hline
\end{tabular}
\end{center}
\caption{Three sets of pseudo-critical temperatures determined from (i) the
chiral susceptibility $\chi^{M_{\rm sub}}_m$, (ii) the
mixed susceptibilities obtained from
temperature, chemical potential
and strange mass derivatives of $M_\ell$,  and (iii) the mixed susceptibility obtained from the temperature derivative of $M$, respectively. 
}
\label{tab:Tpc}
\end{table*}

\subsection{Pseudo-critical temperatures}
\label{sc:pseudo}
As we exploit in our determination of the chiral critical surface,
scaling relations, which are valid at small values of the symmetry
breaking parameter $H$, it is worthwhile to analyze first the
behavior of pseudo-critical temperatures at vanishing chemical
potentials.

\begin{figure*}[t]
\includegraphics[width=0.44\textwidth]{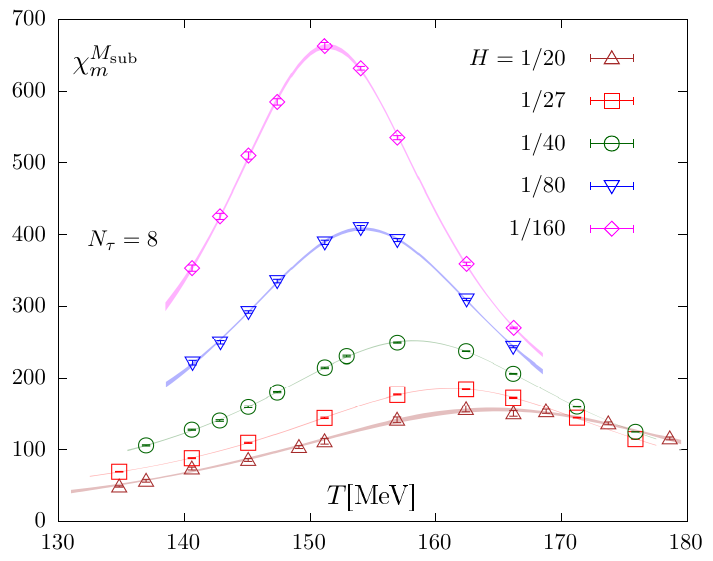}
\includegraphics[width=0.44\textwidth]{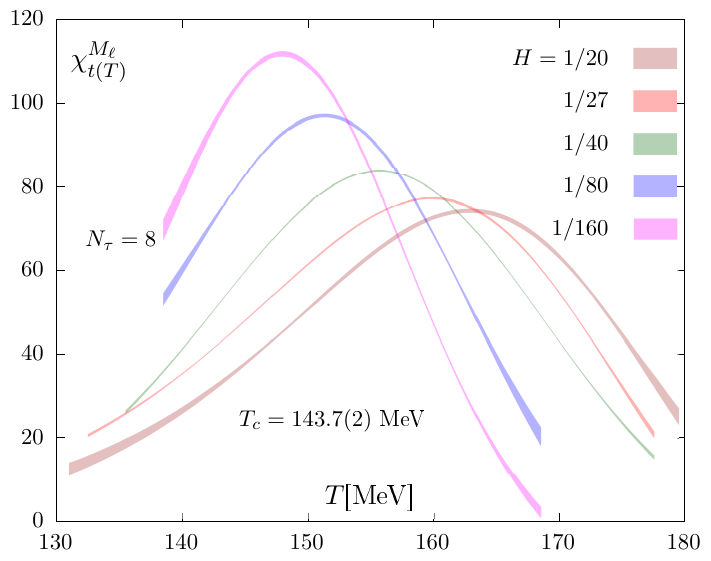}

\includegraphics[width=0.44\textwidth]{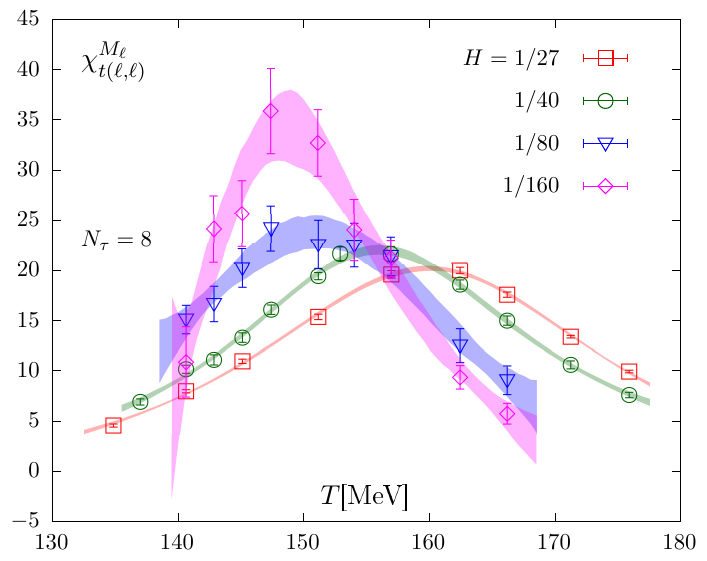}
\includegraphics[width=0.44\textwidth]{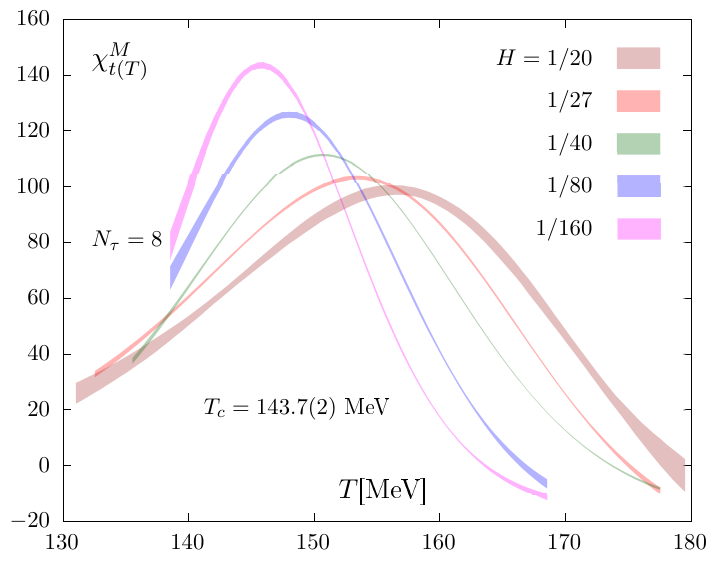}
\caption{The chiral susceptibility,
$\chi^{M_{\rm sub}}_m$ (top, left) and the 
mixed susceptibilities
$\chi^{M_\ell}_{t(T)}$ (top, right), $\chi^{M_\ell}_{t(\ell,\ell)}$
(bottom, left), and $\chi^M_{t(T)}$
(bottom, right),
plotted versus the
temperature for various $H$ values.
Results for the susceptibilities shown on the right hand side have been obtained from derivatives of the rational polynomial fits
to the order parameter data for $M_\ell$ (top) and $M$ (bottom), respectively.
}
\label{fig:chitandchimu}
\end{figure*}

As pointed out in the previous section  pseudo-critical temperatures are not
unique. Using extrema in second  derivatives of the partition function with 
respect to either (i) the external field coupling $H$, or (ii) mixed derivatives with 
respect to $H$ and one of the temperature-like couplings ($T,\ \mu_\ell^2, \mu_\ell\mu_s, \mu_s^2$), or (iii) by
using the temperature derivative of a particular renormalized version of the order parameter,
we define three classes of pseudo-critical 
temperatures\footnote{In general one can also determine a pseudo-critical 
temperature from the susceptibility obtained as second derivative with respect 
to temperature (specific heat). However, in the $O(N)$ universality class, this 
susceptibility does not diverge as the relevant critical exponent $\alpha$ is 
negative.}, which converge for $H\rightarrow 0$ to the uniquely defined 
critical temperature $T_c$. 
In the universal scaling region these three sets of observables
yield three different sets of 
pseudo-critical temperatures, which
will be ordered according to the 
universal position of the extrema
of the relevant scaling functions
given in
Eqs.~\ref{zzp}-\ref{zztM},
{\it i.e.} in the scaling regime we
expect to find
\begin{equation}
    T_{pc,m}(H)>T_{pc,t}(H)>T_{pc,(t,M)}(H)\; .
\end{equation}
In Fig.~\ref{fig:chitandchimu} we show results for the three types
of susceptibilities, the chiral susceptibility ($\chi^{M_{\rm sub}}_m$: top, left), two
versions of mixed susceptibilities obtained from the unrenormalized order parameter $M_\ell$ by taking derivatives with respect to temperature ($\chi^{M_\ell}_{t(T)}$: top, right) or the light quark chemical potential ($\chi^{M_\ell}_{t(\ell,\ell)}$: bottom, left),
and the mixed susceptibility obtained as temperature derivative of the renormalized order parameter $M$ ($\chi^M_{t(T)}$: bottom, right). 

 While the susceptibilities $\chi^{M_{\rm sub}}_m$ and $\chi^{M_\ell}_{t(\ell,\ell)}$ are directly obtained from simulation data,
the susceptibilities $\chi^{M_\ell}_{t(T)}$ and
$\chi^M_{t(T)}$ are obtained by taking
$T$-derivatives of the rational polynomial ans\"atze used to fit $M_\ell$
and $M$, respectively.
As expected from Eqs.~\ref{chitm-scaling}-\ref{chil-scaling}, we find that the rise of the maxima of the
mixed susceptibilities with decreasing $H$ is slower than that of the chiral
susceptibility $\chi^{M_{\rm sub}}_m$.

We also
calculate the mixed susceptibility 
$\chi^{M_\ell}_{t(s,s)}$,
obtained by taking two derivatives of $M_\ell$ with respect to the strange quark chemical potential. This susceptibility
first drops in the large quark mass region and starts to increase only for $H\le 1/80$. It suggests
that regular terms still contribute strongly to this 
susceptibility, which obviously is most sensitive to the strange 
quark sector. In our analysis of pseudo-critical temperatures
we therefore do not include 
$\chi^{M_\ell}_{t(s,s)}$.

\begin{figure}[t]
    \includegraphics[width=0.48\textwidth]{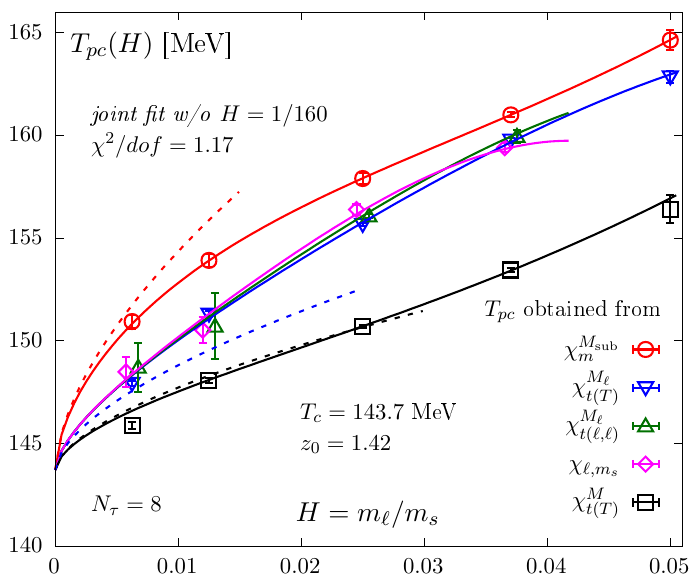}
        \caption{Pseudo-critical temperatures obtained from maxima in (i) $\chi^{M_{\rm sub}}_m$,
(ii) $\chi^{M_\ell}_{t(T)}$, and $\chi^{M_\ell}_{t(\ell,\ell)}$, $\chi_{\ell,m_s}$ and (iii) $\chi^M_{t(T)}$. 
These three sets of observables 
are related to universal maxima
in three different sets of scaling functions (see text).
Lines show joined fits performed 
with a scaling ansatz and including
corrections arising from 
correction-to-scaling as well as regular terms as discussed in the text.
Dashed lines show the leading
$H$-dependent correction arising
from the universal scaling ansatz
in this fit (see text).
	}
\label{fig:Tpc}
\end{figure}

Pseudo-critical temperatures are obtained from maxima of the fit functions used to fit the susceptibilities. For $\chi^{M_\ell}_{t(T)}$ and $\chi^{M}_{t(T)}$\footnote{Note that we used in Eqs.~\ref{chi-t} and \ref{chi-tM}, $T_c$ as a temperature independent normalization for the dimensionless mixed susceptibilities $\chi^{M_\ell}_{t(T)}$ and $\chi^{M}_{t(T)}$. This
insures that the maxima of these mixed
susceptibilities agree
with the inflection points of the order parameters $M_\ell$ and $M$, respectively.} , we use the maxima of the $T$-derivative of the rational polynomial functions used to fit $M_\ell$ and $M$ respectively. The data for $\chi^{M_{\rm sub}}_m$ and $\chi^{M_\ell}_{t(\ell,\ell)}$ have been interpolated directly using rational polynomial ansatze from which the maxima have been calculated. We use [3,2] Pade polynomials for all of the above fits. The error bands have been obtained from a bootstrap analysis.
We summarize results of these fits 
in Table~\ref{tab:Tpc} and Fig.~\ref{fig:Tpc}.
Note that in the table as well as in the figure, we also include results for 
yet another mixed susceptibility
obtained from a derivative of $M_\ell$ with respect to the strange
quark mass. We discuss this observable in more detail in Section VI.

In the case of the mixed susceptibility obtained from $M_\ell$ we actually
use three different observables, obtained by taking derivatives of the order 
parameter with respect to $T$, the chemical potential $\hmu_\ell$ or the strange quark mass $m_s$ (see Section VI for a discussion of the latter).
In the scaling regime all these observables define the pseudo-critical temperature $T_{pc,t}$. 
In fact, as can be seen in 
Fig.~\ref{fig:Tpc} the location of maxima of these mixed susceptibilities is 
identical within errors even for the physical 
value of the light to strange quark mass ratio, $H=1/27$.
In addition to the mixed susceptibility $\chi^{M_\ell}_{t(T)}$, defined as the $T$-derivative
of $M_\ell$, we also use the 
mixed susceptibility $\chi^M_{t(T)}$,
which  is obtained as the $T$-derivative of the
renormalized order parameter $M$.
This defines another
pseudo-critical temperature, $T_{pc,(t,M)}$.

As can be seen in Fig.~\ref{fig:Tpc} the pseudo-critical temperatures
reflect the ordering of the 
universal scaling relations
for all $H\le 1/20$.
The pseudo-critical temperatures,
however, are generally located outside the 
temperature range in which we found good scaling behavior for the
order parameter $M$, shown in Fig.~\ref{fig:Sigma}. 
When fitting data for the pseudo-critical temperatures to extract $T_c$ we
thus need to take into account
also the influence of sub-leading
corrections to the location of the
maxima in susceptibilities.

We have fitted the data shown in Fig.~\ref{fig:Tpc} using  
scaling ans\"atze appropriate for 
the different observables used to
define a pseudo-critical temperature and allowing for contributions from corrections-to-scaling as well as regular terms. We start with an ansatz for the 
unrenormalized order parameter $M_\ell$,
\begin{eqnarray}
    M_\ell&=& h_0^{-1/\delta} H^{1/\delta} \left( f_G(z) + c H^{\omega\nu_c}{f}_{G,cts}(z)+{\cal O}(H^{2\omega\nu_c})\right) \nonumber \\
    &&+ H
    \sum_{n=0}^{n_{max}} a_n t^n 
+{\cal O}(H^3)
\; ,
\label{fit-ansatz}
\end{eqnarray}
and similarly for the renormalized
order parameter $M$, where $f_G(z)$
gets replaced by $f_G(z)-f_\chi(z)$. The definition of the  corrections-to-scaling function $f_{G,cts}$ and further details about the above ansatz are provided in Appendix \ref{app:cts}. 
Calculating the various susceptibilities by starting from Eq.~\ref{fit-ansatz} one obtains
the influence of the sub-leading terms on
the locations of 
the maxima of the susceptibilities by expanding the order parameter ansatz for small values of $H$ \cite{Bazavov:2011nk}
in the vicinity of the relevant, universal peak locations $z_m$, $z_t$ or
$z_{t,M}$.
Keeping terms up to $n_{max}=2$
for the determination of $T_{pc,m}$
and $n_{max}=3$
for the determination of $T_{pc,x}$,
$x=t, (t,M)$, insures that in
both cases regular terms linear in $t$ are kept in the determination
of maxima of the susceptibilites. 
This gives for the position of 
a peak in the mixed susceptibilities
\begin{eqnarray}
T_{pc,x}(H) &=& 
T_c \left( 1 + t_{1,x} H^{1/\beta\delta}
+ t_{c,x} H^{1/\beta\delta +\omega\nu_c}
\right. \nonumber \\
&&\left. 
	+t_{2,x} H^{1+(3-\beta)/\beta\delta} +t_{3,x} H^{1+(4-\beta)/\beta\delta}
\right)  \nonumber \\
&&\hspace{2.0cm} x=t,\ (t,M) \; , \label{mixedTpc}
\end{eqnarray}
whereas for the chiral susceptibility one has \cite{Bazavov:2011nk},
\begin{eqnarray}
\label{chiralTpc}
T_{pc,m}(H) &=& 
T_c \left( 1 + t_{1,m} H^{1/\beta\delta}
+ t_{c,m} H^{1/\beta\delta +\omega\nu_c}
\right.  \\
&&\left. 
	+t_{2,m} H^{1+(2-\beta)/\beta\delta} +t_{3,m} H^{1+(3-\beta)/\beta\delta}
\right) \; . \nonumber
\end{eqnarray}
The coefficient $t_{1,x}$ of the leading $H$-dependent correction to $T_c$ is related to the universal
parameters $z_x$, given in
Eqs.~\ref{zzp}-\ref{zztM},
and the non-universal scale $z_0$,
\begin{equation}
    t_{1,x}\equiv \frac{z_x}{z_0}\; ,\; x=m,\ t,\ (t,M)\; .
    \label{t1x}
\end{equation}
We performed fits in which $T_c$ and 
$t_{1,x}$ are kept as free fit parameters as well as using for $T_c$ and $t_{1,x}$
the values determined from the scaling fits to the order parameter $M$ given in Eqs.~\ref{Tcfix}, \ref{z0fix}.
Using  $T_c$ and $t_{1,x}$ as free fit parameters gives result that are 
consistent with Eqs.~\ref{Tcfix}, \ref{z0fix}.

We performed joint fits to data
for pseudo-critical temperatures obtained from five 
different susceptibilities,
demanding that they yield the
same critical temperature in the chiral limit. 
It turns out that in the case of the mixed susceptibility
$\chi^M_{t(T)}$
these fits are not sensitive to a term proportional to $t_{3,(t,M)}$.
We thus set
$t_{3,(t,M)}=0$ in our final fits.
Furthermore, we need to control the
influence of 
the smaller volume used in our
determination of $T_{pc}$ in
calculations with $H=1/160$. We
therefore performed fits (i) leaving out or (ii) including the data for $H=1/160$ as well as (iii) using the data for $H=1/160$ 
where we corrected the $T_{pc}$
values, obtained for $H=1/160$, by
a global shift of $0.25$~MeV. This is
in accordance with the finite volume 
dependence found in \cite{Ding:2019prx}. 
For $N_\tau=8$, $H=1/80$ it was found there that pseudo-critical temperatures 
determined at $z_{L,b}\simeq 1.1$ and
$0.8$ differ by about $0.25$~MeV.
We use this value also as an estimate for finite volume effects in our data 
obtained for $H=1/160$ at $z_{L,b}\simeq 1.1$.

We find that fits using these three different
approaches yield similar results 
for the chiral transition temperature
$T_c$,
\begin{equation}
    \hspace*{-0.2cm}T_c =
    \begin{cases}
   143.9(5)~{\rm MeV}, & {\rm without}~ H=1/160 ~{\rm data}\\
   142.8(3)~{\rm MeV}, & {\rm with}~ H=1/160 ~{\rm data} \\
   143.4(3)~{\rm MeV}, & {\rm vol.~ corrected}~ H=1/160 ~{\rm data}
    \end{cases}
\end{equation}
The resulting critical temperatures,
extracted from the fits to pseudo-critical temperatures,
thus are in good agreement with that determined from
the scaling fit to the order parameter $M$.
We also find that keeping the slope
parameters $t_{1,x}$ as free fit parameters gives results that are 
consistent with those obtained by fixing these coefficients to the universal values given in Eq.~\ref{t1x}. 

For our final fits we
then fixed $T_c$ and $t_{1,x}$ using
for $(T_c,z_0)$ the values given in Eqs.~\ref{Tcfix}, \ref{z0fix} and the
universal parameters $z_m$, $z_t$, $z_{t,M}$, given in
Eqs.~\ref{zzp}-\ref{zztM}. Our 
final joint fit of all $T_{pc}(H)$
data sets shown in Fig.~\ref{fig:Tpc}
thus only depends on fit parameters
fixing contributions from regular
and correction-to-scaling terms.

The fit result, obtained by leaving out data obtained for $H=1/160$, is shown in Fig.~\ref{fig:Tpc}. This fit
has a $\chi^2/dof=1.17$. 
The contribution arising only 
from the universal part of this 
fit, $T_{pc,x}=T_c \left( 1 + t_{1,x} H^{1/\beta\delta}\right)$,
is shown by dashed lines in  Fig.~\ref{fig:Tpc}. 
As can be seen 
this universal part dominates the fit of 
pseudo-critical temperatures, obtained from the mixed susceptibility $\chi_{t(T)}^{M}$, at least for $H\le 1/40$. In the case of 
$\chi_{t(T)}^{M_\ell}$ or $\chi_m^{M_{\rm sub}}$, however,
contributions from sub-leading corrections become important already for $H> 1/80$. This is in accordance with the conclusion drawn from the scaling fit performed for the order parameter $M$, {\it i.e.} sub-leading corrections need to be 
taken into account for temperatures larger than $T\simeq 148~{\rm MeV}$.

 \section{Curvature coefficients and pseudo-critical temperatures at \boldmath $\vec{\mu}\ne 0$}
 \label{sec:curve}
 We have introduced the dependence of pseudo-critical temperatures on light and strange  quark chemical potentials in Eq.~\ref{Tpcx}.
 Equivalently this can be 
 written in the $(B, S)$ basis
 where the chemical potentials
 $\mu_B$ and $\mu_S$ are obtained from those in the flavor basis using
  \begin{equation}
   \mu_\ell = \frac{1}{3}\mu_B, \quad \mu_s = \frac{1}{3}\mu_B - \mu_S \; .
   \label{muBSls}
  \end{equation}

 With this 
 we may rewrite Eq.~\ref{Tpcx} as,
  \begin{eqnarray}
     T_{pc,x}(\hmu_B,\hmu_S,H) &=& 
     T_c \Bigl( 1 - (\kappa_2^B \hmu_B^2
  + \kappa_2^S \hmu_S^2 + 2\kappa_{11}^{BS} \hmu_B \hmu_S)
   \nonumber \\
     && + \frac{z_{x}}{z_0}H^{1/\beta\delta} \;\; \Bigr)\; ,\; x= m,\ t,\ (t,M) \; ,
 \label{TpcBS}
 \end{eqnarray}
 where the curvature coefficients
 calculated for vanishing chemical potentials using the quark-flavor or
 conserved charge basis, respectively, are related to each other through,
 \begin{eqnarray}
     \kappa_2^B &=&\frac{1}{9}\left( \kappa_2^\ell+2 \kappa_{11}^{\ell s} + \kappa_2^s \right) 
     \label{k2B} \\
     &=& \frac{\kappa_2^\ell}{9}\left( 1 +2\frac{\kappa_{11}^{\ell s}}{\kappa_2^\ell} +
     \frac{\kappa_2^{s}}{\kappa_2^\ell} \right) \; , \nonumber \\
     \kappa_2^S &=& \kappa_2^s \; ,
     \label{k2S} \\
     \kappa_{11}^{BS} &=& -\frac{1}{3}\left(  \kappa_2^s +\kappa_{11}^{\ell s}\right) \\
     &=& -\frac{\kappa_2^s}{3}\left( 1  +\frac{\kappa_{11}^{\ell s}}{\kappa_2^s}\right)
     \; . \nonumber
     \label{k11BS}
 \end{eqnarray}
We also note that the mixed curvature coefficients 
$\kappa_{11}^{\ell s}$ and $\kappa_{11}^{BS}$
 carry quite different information. While the former is non-zero only due to flavor correlations arising, for instance, in high 
 temperature perturbation theory only at ${\cal O}(g^6\ln(1/g)$ \cite{Blaizot:2001vr}, the latter receives contributions also
 from diagonal terms in the strangeness sector.
 
 The pseudo-critical temperatures
in the $(B, S)$ basis 
(Eq.~\ref{TpcBS}) may be 
 re-written when imposing constraints on the strangeness chemical potential, e.g. in the
 case of a strangeness neutral
 medium.  
 To leading order the strangeness chemical 
 potential, insuring vanishing net 
 strangeness number in an isospin symmetric medium is given by
 $\hmu_S=s_1(T,H) \hmu_B$ \cite{Bazavov:2017dus}, with
 $s_1(T,H)=-\chi_{11}^{BS}/\chi_2^S$. 
 With this we  obtain for the $\hmu_B$-dependence of
 pseudo-critical temperatures in strangeness-neutral, isospin-symmetric  ($\mu_Q=0\Leftrightarrow \mu_I=0$) matter,
\begin{eqnarray}
     T_{pc,x}^{n_S=0}(\hmu_B, H) &=& 
     T_c \Bigg( 1 
     - \Big(\kappa_2^B+ s_1^2(T_c,0) \kappa_2^S \nonumber \\
     &&+ 2 s_1(T_c,0)\kappa_{11}^{BS}\Big) \hmu_B^2
     + \frac{z_{x}}{z_0}H^{1/\beta\delta}
     \Bigg)\; ,
      \nonumber \\
     && \hspace{1.8cm} \;\;\;\; x= m,\ t,\ (t,M)\; .
\label{Tpc-neutral}
\end{eqnarray}     
A recent calculation of $s_1(T,H)$ for physical
values of the light and strange quark masses is
given in \cite{Bollweg:2021vqf}.
This yields $s_1(T_{pc},1/27)\simeq 0.236(5)(6)$. In Appendix~\ref{app:s1}
we present results for the dependence of $s_1(T,H)$ on the quark mass ratio $H$. An extrapolation to 
the chiral limit yields
$s_1(T_c,0)=0.216(3)$.
 
 A calculation of curvature coefficients
 making use of the universal scaling relations
 discussed in the previous section has been
 performed previously at physical values
 of the quark masses \cite{Kaczmarek:2011zz}. 
 Most calculations, however, make use of 
 Taylor expansions for the dependence of the maximum of the
 chiral susceptibility $\chi^{M_{\rm sub}}_m$
 or inflection points of the chiral
 condensates. In that case one starts
 from either Taylor expansions of the 
 observables $\Oc\equiv M,\ \chi_m,\ \chi_t $\cite{Kaczmarek:2011zz,Bazavov:2018mes,Bonati:2018nut} 
 or simulations performed with
 imaginary values of chemical potentials
 \cite{Bonati:2015bha,Bellwied:2015rza,Borsanyi:2020fev}, 
 \begin{eqnarray}
     \Oc(T,\mu_\ell,\mu_s) &=& \Oc(T_{\Oc,m},0,0) +\frac{\partial \Oc}{\partial T}\left( T-T_{\Oc,m} \right) \nonumber \\
     &&+\frac{1}{2} 
     \frac{\partial^2 \Oc}{\partial \hmu_\ell^2} \hmu_\ell^2
     +\frac{1}{2} 
     \frac{\partial^2 \Oc}{\partial \hmu_s^2} \hmu_s^2
      \label{Otaylor} \\
     &&+\frac{\partial^2 \Oc}{\partial \hmu_\ell\partial \hmu_s} \hmu_\ell\hmu_s +\Oc(\hmu^4,\Delta T \hmu^2,(\Delta T)^2)\; . \;
     \nonumber
 \end{eqnarray}
Here $\Delta T=(T-T_{\Oc,m})/T_{\Oc,m}$ and
$T_{\Oc,m}\equiv T_{\Oc,pc}(m_\ell)$ denotes the 
pseudo-critical temperature, which 
 at non-vanishing values of light quark 
 masses $m_\ell$ depends on the particular observable $\Oc$ used for
 its definition. As discussed in the previous section, in the chiral limit all these pseudo-critical temperatures will converge to the uniquely defined phase transition temperature $T_c(\vec{\mu})$ also for non-vanishing chemical potentials.
 
 \begin{figure*}
    \includegraphics[width=0.32\textwidth]{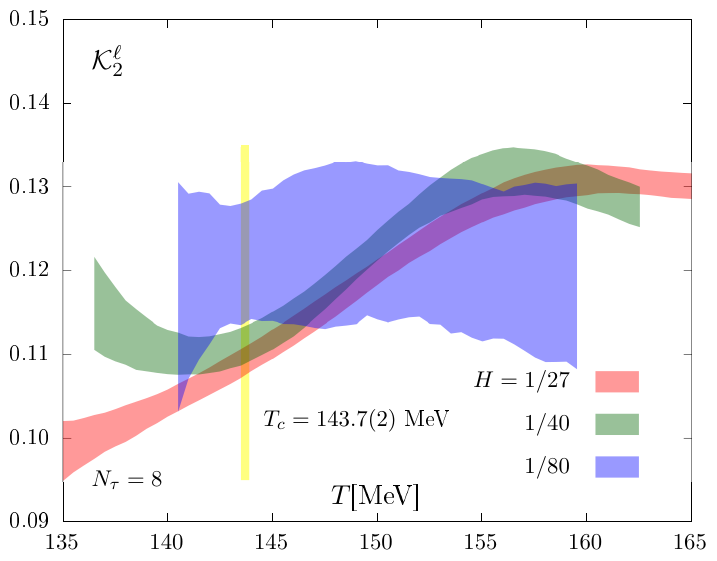}
    \includegraphics[width=0.32\textwidth]{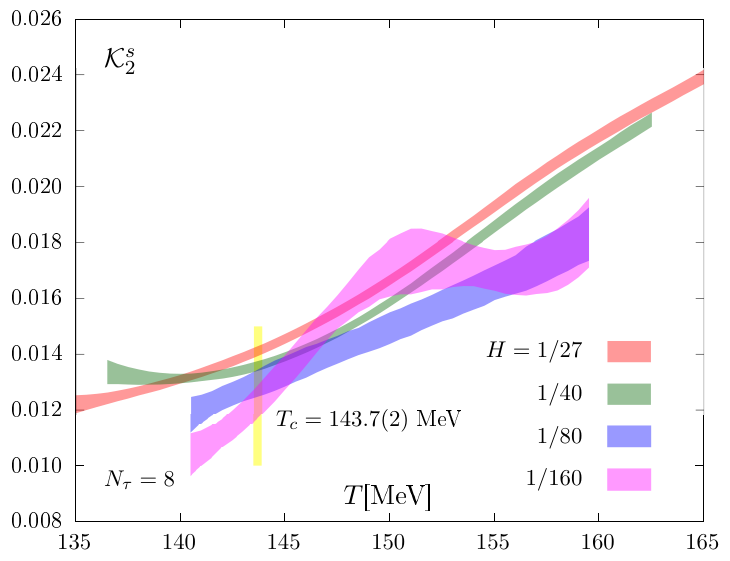}
    \includegraphics[width=0.32\textwidth]{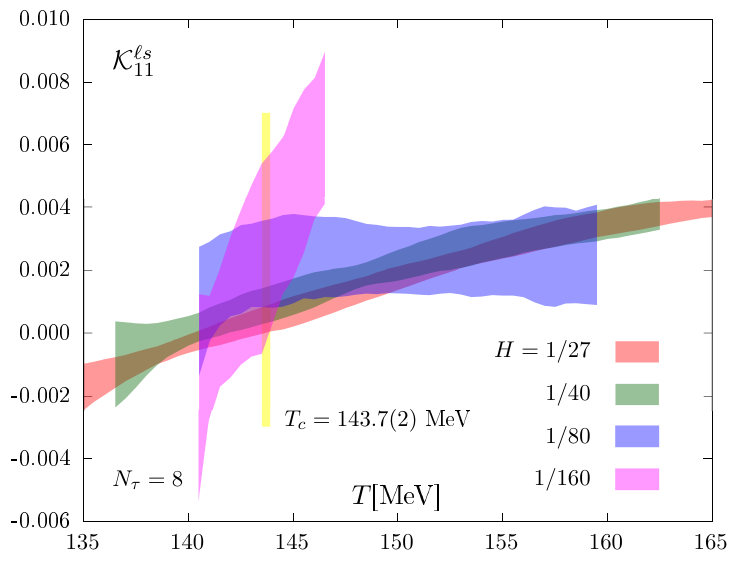}
\caption{Curvature coefficients determined from $M_\ell$ using the 
universal scaling ansatz for $M_\ell$.
Shown are ${\cal K}_2^l$ (left) and ${\cal K}_2^s$ (middle), and  
${\cal K}_{11}^{ls}$ (right) as function
of temperature. The final results for the
curvature coefficients are determined
at $T_c$ (yellow band)
as defined in Eqs.\ref{kappaf} and \ref{kappals}.
}
\label{fig:kappa-flavor}
\end{figure*}
 
\subsection{Curvature coefficients from the order parameter}

Although the curvature coefficients
are defined at $T_c$, we may
evaluate the relevant observables at
any value of the temperature $T$
and for $\hmu_\ell=\hmu_s=0$. 
Making use of the fact that derivatives
of the order parameter with respect to 
energy-like couplings, {\it i.e.} temperature or chemical potentials, will be dominated by the diverging singular contribution, and using
the Taylor expansion of the energy-like
scaling field $u_t$, we
determine the curvature coefficient
from a ratio of derivatives of
order parameter $M_\ell$ with 
respect to chemical potentials and 
temperature, respectively,
\begin{eqnarray}
    {\cal K}_2^f(T,H) &=& \frac{1}{2T_c}\left(
    \frac{\partial^2 M_\ell/\partial \hmu_f^2}{\partial M_\ell/\partial T}\right)_{(T,\vec{\mu}=0)}  ,\;
    f=\ell,\ s \; ,
    \label{kappa-scaling-l2} \\
    {\cal K}_{11}^{\ell s}(T,H) &=& \frac{1}{2T_c}\left(
    \frac{\partial^2 M_\ell/\partial \hmu_\ell \partial \hmu_s}{\partial M_\ell/\partial T}\right)_{(T,\vec{\mu}=0)} \;\: . \;\;
    \label{kappa-scaling-ls}
\end{eqnarray}
The curvature coefficients are 
then given by 
\begin{eqnarray}
\kappa_2^f &\equiv& {\cal K}_2^f(T_c,0) \;\: ,\;\; 
    f=\ell,\ s \; ,
    \label{kappaf}\\
\kappa_{11}^{\ell s} &\equiv& {\cal K}_{11}^{\ell s}(T_c,0) \; .
\label{kappals}
\end{eqnarray}

In the vicinity of the critical point at
$T_c$ and for small values of the 
chemical potentials  we may use
the scaling ansatz for the order parameter $M_\ell$, given
in Eq.-\ref{M-scaling}, to determine the curvature coefficients in the flavor basis.
We will also take into account
possible regular contributions
to the order parameter using a linear
ansatz similar to Eq.-\ref{fit-ansatz},
\begin{eqnarray}
    f_r(T,\hmu_\ell,\hmu_s) &=& H\Big( a_0+a_t \frac{T-T_c}{T_c} 
    \nonumber \\
    &&+a_\ell \hmu_\ell^2 +
    a_s \hmu_s^2 +2 a_{\ell s} \hmu_\ell \hmu_s
    \Big) \; .
    \label{regular}
\end{eqnarray}
Using the expansion of $f'_G(z)$
for small $z$, {\it i.e.} in the vicinity of $T_c$
\begin{eqnarray}
 f'_G(z)&=& f'_G(0) + f''_G(0) z
 + {\cal O}(z^2)
 \nonumber \\
 &=& f'_G(0) + z_0 f''_G(0) H^{-1/\beta\delta} t + {\cal O}(t^2/H^{2/\beta\delta}) \; ,
\end{eqnarray}
we may expand ${\cal K}_2^f(T,H)$
in the vicinity of $T_c$,
\begin{eqnarray}
{\cal K}_2^f(T,H)
&=& \frac{\kappa_2^f h^{(\beta-1)/\beta\delta} f'_G(z)/t_0 + H a_f}{h^{(\beta-1)/\beta\delta} f'_G(z)/t_0 + H a_t }
    \nonumber \\
    &=& \frac{\kappa_2^f + \tilde{a}_f  H^{1+(1-\beta)/\beta\delta}/f'_G(z)}{1+\tilde{a}_t H^{1+(1-\beta)/\beta\delta}/f'_G(z)} 
        \label{kapparatio}  \\
&=& \kappa_2^f +A_f H^{1+(1-\beta)/\beta\delta}+ B_f H^{1-1/\delta} t \; ,
    \nonumber \\
    &&\hspace{4.5cm} f=\ell,\ s\; ,
    \nonumber
\end{eqnarray}
where the last line of Eq.~\ref{kapparatio} gives explicitly the leading quark mass 
and temperature dependence arising from the rescaled regular term.
Similar relations hold for the 
flavor off-diagonal curvature term
$\kappa_{11}^{\ell s}$. Furthermore,
we may directly calculate ratios
of curvature coefficients, e.g.
\begin{eqnarray}
\frac{\kappa_s}{\kappa_\ell}
&=&\lim_{m_\ell\rightarrow 0}
\left.\frac{{\cal K}_2^s(T,H)}{{\cal K}_2^\ell(T,H)}\right|_{(T_c,\vec{\mu}=0)}
\nonumber \\
&=&
\lim_{m_\ell\rightarrow 0}
\left. \frac{\partial^2 M_\ell /\partial\hmu_s^2}{\partial^2 M_\ell/\partial \hmu_\ell^2}\right|_{(T_c,\vec{\mu}=0)}
\; .
\label{ratio-ls}
\end{eqnarray}
We note that the scaling function
$f'_G(z)$ as well as the 
critical exponents entering the scaling relations, of course, are
specific to a given universality class. However, the basic definition
of the curvature coefficients,
given in  Eqs.~\ref{kappa-scaling-l2} and \ref{kappa-scaling-ls}
as well as Eq.~\ref{ratio-ls}, do neither 
depend on the value of these exponents nor do
they depend on the specific form of the 
scaling functions. The fact, that the global
symmetry group for staggered fermions, used 
in our calculations at finite lattice spacings, is different from that of QCD in the continuum
limit thus does not enter in these relations.

Results for ${\cal K}_2^f(T,H)$ and
${\cal K}_{11}^{\ell s}(T,H)$  are
shown in Fig.~\ref{fig:kappa-flavor}. 
From this we determine the  
curvature coefficients in flavor-basis at $T_c$. It is 
apparent from Fig.~\ref{fig:kappa-flavor} that the 
curvature coefficient related to 
the light quark chemical potential
is almost an order of magnitude larger than that in the direction of the strange quark chemical potential. The off-diagonal curvature coefficient, in turn, is 
an order of magnitude smaller than 
$\kappa_2^s$. We give results
for the various curvature coefficients and different 
quark mass ratios $H$ in Table-\ref{tab:curvature}.

\begin{table*}[htb]
\begin{center}
\begin{tabular}{|c||c|c|c||c|c||c|}
\hline
$H$ & $\kappa_2^\ell$& $\kappa_2^s$& $\kappa_{11}^{\ell s}$ & $\kappa_2^B$
& $\kappa_{11}^{BS}$ &$\kappa^{m_s}$ \\
\hline 
1/27  & 0.109(2) & 0.0141(2) & 0.0004(4) & 0.0138(2) & -0.0048(1) & 0.147(1) \\
1/40  & 0.111(2) & 0.0136(2) & 0.0009(5) & 0.0141(3) & -0.0048(2) & 0.129(1) \\
1/80  & 0.121(7) & 0.0129(5) & 0.002(1)  & 0.015(1)  & -0.0052(5) & 0.111(2) \\
1/160 & 0.12(1)  & 0.0123(6) & 0.003(3)  & 0.015(2)  & -0.0048(10)  & 0.098(2) \\
\hline
0 & 0.122(7) & 0.0124(5) & 0.003(2) & 0.015(1) & -0.0050(7) & 0.097(2)\\
\hline
\end{tabular}
\end{center}
\caption{Curvature coefficients in the flavor and conserved charge basis, respectively. Given are results for several values of the quark mass ratio $H$.  In the last column the curvature coefficient defining the dependence of $T_c$ on the strange quark mass in the vicinity of the physical strange quark mass value is given. The last row gives our estimates for the chiral limit extrapolated values based on fits using in addition to the 
scaling ansatz a regular term proportional to $H$.
}
\label{tab:curvature}
\end{table*}

\begin{figure}
\centering
\includegraphics[width=0.45\textwidth]{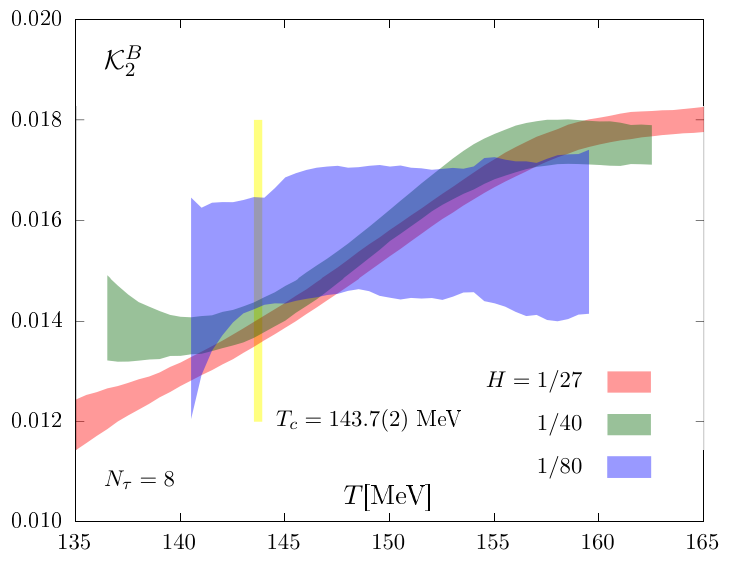}
\includegraphics[width=0.45\textwidth]{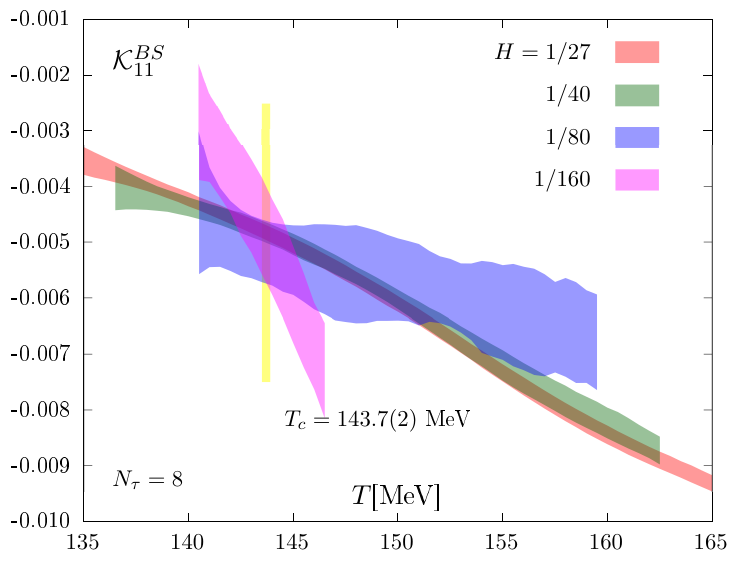}
\caption{The ratios ${\cal K}_2^B$ (top)
and ${\cal K}_{11}^{BS}$ (bottom) defined 
in analogy to the corresponding functions
in the flavor basis, ${\cal K}_2^\ell$
(Eq.~\ref{kappaf}) and ${\cal K}_{11}^{\ell s}$ (Eq.~\ref{kappals}), respectively. The chiral extrapolation
at $T_c$ defines the
curvature coefficients $\kappa_2^B$ (top) and $\kappa_{11}^{BS}$ (bottom).}
\label{fig:kappaB}
\end{figure}

The resulting curvature coefficients in the ($B, S$) basis can be obtained by using
similar equations corresponding to Eqs.~\ref{kappa-scaling-l2} and \ref{kappa-scaling-ls}.
The curvature coefficients
$\kappa_2^B$ and $\kappa_{11}^{BS}$
are shown in Fig.~\ref{fig:kappaB}.
Note that $\kappa_2^S=\kappa_2^s$ is shown already in Fig.~\ref{fig:kappa-flavor}.
Moreover, it is evident from
Fig.~\ref{fig:kappaB} that
the off-diagonal coefficient
$\kappa_{11}^{BS}$ is clearly
non-zero and negative. As 
expected, its value is dominated
by the contribution of the diagonal curvature coefficient 
$\kappa_2^S$.

We fitted the results for
curvature coefficients in the flavor basis, ($\kappa_2^\ell$, $\kappa_2^s$, $\kappa_{11}^{\ell s}$), as well as in
the conserved charge basis,
($\kappa_2^B$, $\kappa_{11}^{BS}$) 
using for the quark mass dependence the 
ansatz given in Eq.~\ref{kapparatio}.
Results of these fits are given in the 
last row of Table~\ref{tab:curvature}.

\begin{figure}
\centering
\includegraphics[width=0.48\textwidth]{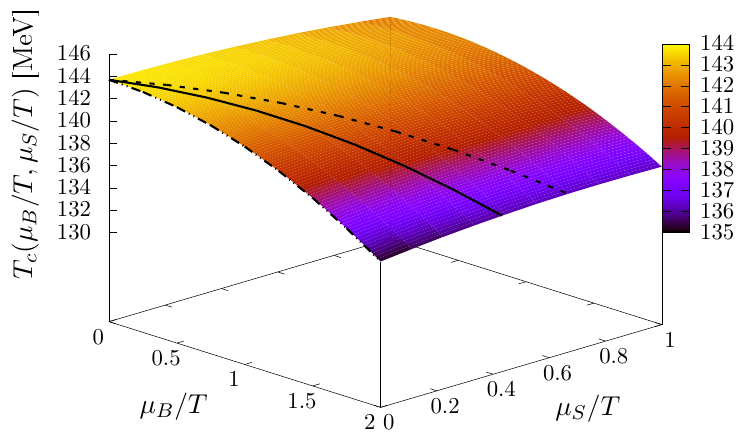}
\caption{Surface of critical 
temperature in the $\mu_B$-$\mu_S$
plane. The solid line shows $T_c$
as function of $\mu_B$ in strangeness neutral matter. The dashed line in the front corresponds to $\mu_S=0$ whereas 
the dashed line at large $\mu_S$ corresponds to the case $\mu_s=0$, which
is equivalent to $\mu_S=\mu_B/3$.
}
\label{fig:Tcsurface}
\end{figure}

With this we find for the curvature 
coefficient in the case of (i) vanishing strangeness
chemical potential, $\mu_S=0 \leftrightarrow \mu_s=\mu_\ell$,
(ii) along the strangeness
neutral line, $n_S=0\leftrightarrow \mu_S=s_1(T,H)\mu_B$, and (iii) 
for vanishing strange quark chemical potential, $\mu_s=0\leftrightarrow \mu_S=\mu_B/3$,
\begin{eqnarray}
    \kappa_2^{B,\mu_S=0} &\equiv&
    \kappa_2^B =
0.015(1) 
    \; ,\\
    \kappa_2^{B,n_S=0} &=&
    \kappa_2^B+ s_1^2(T_c,0) \kappa_2^S + 2 s_1(T_c,0)\kappa_{11}^{BS}
    \nonumber \\
    &=& 0.893(35)\ \kappa_2^B 
        \; , \\
            \kappa_2^{B,\mu_s=0} &=&
    \frac{1}{9} \kappa_2^\ell=
    0.968(23)\ \kappa_2^{B,n_S=0} \; .
\end{eqnarray}
We thus find that the curvature coefficient
in strangeness neutral matter 
is about 10\% smaller than that for $\mu_S=0$.
In the curvature coefficient for the case $\mu_s=0$, which corresponds to the relation $\mu_S=\mu_B/3$ and which 
sometimes is used as convenient approximation for the strangeness neutral case, 
is only about 3\% smaller than
$\kappa_2^{B,n_S=0}$.
The above results for the curvature coefficients are reflected in the critical temperature surface in the $\hmu_B$-$\hmu_S$ plane shown 
in Fig.~\ref{fig:Tcsurface}.
In that figure a solid line is 
shown for the case that corresponds to the strangeness neutral line in the $\hmu_B$-$\hmu_S$ plane and two 
dashed lines show lines for constant $\mu_S=0$ and $\mu_s=0$,
respectively.

The results obtained here agree within
errors with determinations of curvature coefficients obtained with physical light and strange quark masses and
various choices for the chemical potentials (for recent summaries see
e.g. \cite{DElia:2018fjp,Borsanyi:2020fev}).
By making use of relations between
curvature coefficients in the $(B,S)$
and ($\ell, s)$ basis, respectively, we
could establish the systematic differences of curvature coefficients 
obtained along different lines of 
fixed $\hmu_S/\hmu_B$.

\section{Dependence of the chiral transition temperature on the strange quark mass}

The dependence of the chiral 
phase transition temperature 
on chemical potentials, determined in the previous section to leading order in
the chemical potentials, has been performed at fixed value 
of the strange quark mass. This has been tuned to its physical
value $m_s^{phy}$. We may also use the various observables, obtained in the course of this 
analysis, to calculate the 
dependence of the chiral phase
transition temperature on the
value of the strange quark mass.

Varying the strange quark mass
from $m_s=0$ to $m_s=\infty$ at 
vanishing values of the two
degenerate light quark masses interpolates between the chiral
limits of 3 and 2-flavor QCD, respectively. Several lattice
QCD calculations, using different discretization schemes, suggest that the chiral
phase transition in both limits
remains to be second order. In
particular, no indication for 
a first order transition in 3-flavor QCD has been found.
We thus expect that universal features of the chiral
phase transition in (2+1)-flavor
QCD is described by the same 
universality class for all finite, non-zero values of the strange quark mass. As the
strange quark mass does not break chiral symmetry in the light quark sector explicitly, it will appear as an external parameter in the energy-like scaling variable $t$, just like the chemical potentials.

We may expand the chiral
phase transition temperature $T_c$, appearing in the definition of the energy-like scaling field $t$ given in Eq.~\ref{coupling-t}, in terms of
a Taylor series around the physical strange mass, $m_s^{phy}$. To leading order
we obtain,
\begin{eqnarray}
    T_c(m_s) &=& T_c(m_s^{phy}) +
    \left. \frac{\partial T_c(m_s)}{\partial m_s}
    \right|_{m_s^{phy}}
    \left(m_s-m_s^{phy}\right)
    \nonumber \\
    &&+{\cal O}((\Delta m_s)^2)\; ,
\end{eqnarray}
Omitting just for clarity the 
$\mu$-dependence of the scaling 
variable $t$, introduced in Eq.~\ref{coupling-t}, we rewrite $t$ as
\begin{eqnarray}
t&=& \frac{1}{t_0}\left[\frac{T}{T_c(m_s)}-1\right] \nonumber\\
&=& \frac{1}{t_0}\left[\left(\frac{T}{T_c(m^{phy}_s)}-1\right) - \kappa^{m_s}\frac{m_s - m_s^{phy}}{m_s^{phy}}\right] \nonumber
\\
&&+{\cal O}((\Delta T)^2,
\Delta T \Delta m_s, (\Delta m_s)^2) 
\end{eqnarray}
where 
\begin{equation}
\kappa^{m_s} = - \frac{ m_s^{phy}}{T_c(m_s^{phy})}\ 
\frac{\partial T_c(m_s)}{\partial m_s}\Big|_{m_s^{phy}} 
\; .
\end{equation}
 
\begin{figure}[t]
\centering
\includegraphics[width=0.46\textwidth]{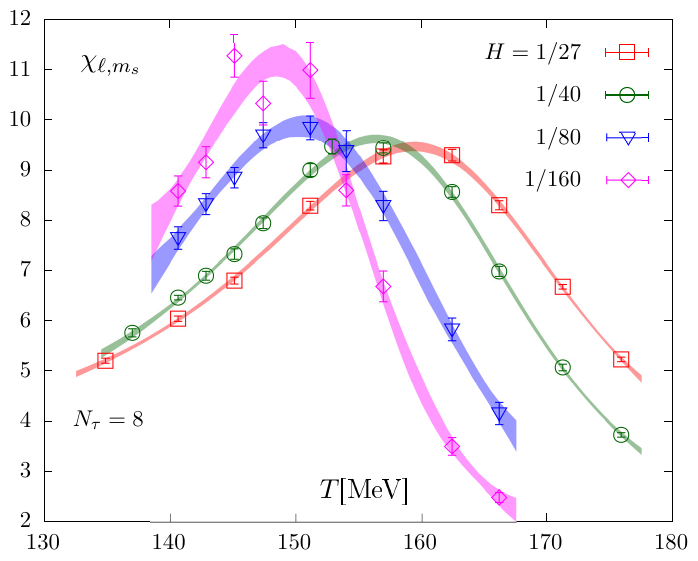}
\includegraphics[width=0.48\textwidth]{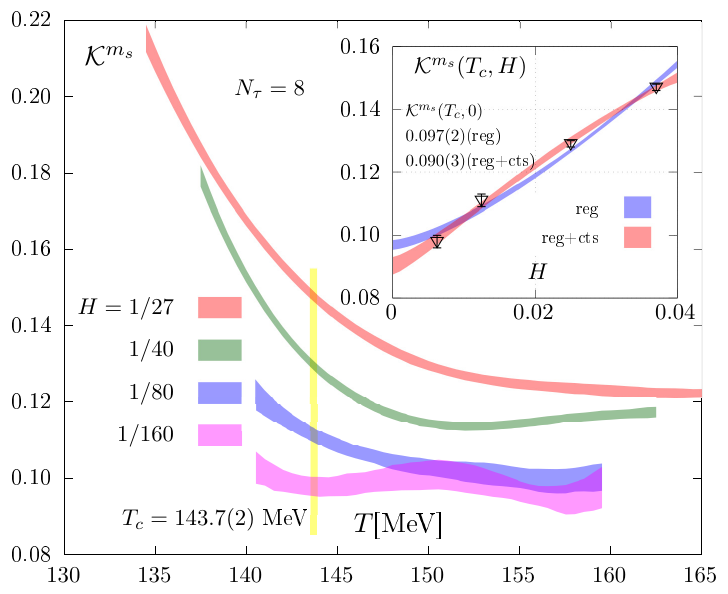}

\caption{Derivatives of light quark condensate with respect to the strange quark mass (top) and
the curvature coefficient in the strange quark mass direction (bottom) obtained from Eq.~\ref{ms-curvature}.
The yellow band shows $T_c$ as given in Eq.~\ref{Tcfix}. In the inset we show
fits using the universal scaling ansatz
and a regular term (reg) only or including 
in addition also the leading
correction-to-scaling term (reg+cts).
}
\label{fig:msder_fK}
\end{figure}

Using the definition of the 
light-strange susceptibility,
introduced in Eq.~\ref{chils},
we may obtain the curvature
coefficient $\kappa^{m_s}$ from
the ratio
\begin{eqnarray}
{\cal K}^{m_s}(T,H) &=& 
\frac{\chi_{\ell, m_s}}{\chi^{M_\ell}_{t(T)}}
\;\: .
\label{ms-curvature}
\end{eqnarray}
Results for the light-strange
susceptibility are shown in Fig.~\ref{fig:msder_fK}~(top).
We see that the peak position
of the susceptibility shifts to 
smaller values as $H$ decreases.
Results for the pseudo-critical temperatures corresponding to the location of these peaks are
also given in Table~\ref{tab:Tpc} and shown in
Fig.~\ref{fig:Tpc}. It is apparent that these pseudo-critical temperatures 
vary with $H$ just like the other mixed susceptibilities
obtained from derivatives of the chiral order parameter $M_\ell$
with respect to a energy-like variable. This 
confirms our expectation that
that the strange quark mass enters the universal scaling relations just like other energy-like couplings.

Results for ${\cal K}^{m_s}(T,H)$, obtained for several values of $H$, are shown
in Fig.~\ref{fig:msder_fK}~(bottom). From this we obtain the 
curvature coefficient $\kappa^{m_s}$ in the chiral limit as
\begin{equation}
    \kappa^{m_s} =\lim_{H\rightarrow 0}
    {\cal K}^{m_s}(T_c,H) \; .
\end{equation}
From a fit to the data for
${\cal K}^{m_s}(T_c,H)$, 
using only the leading correction
arising from a regular term (reg) proportional to $H$ as well as the
additional sub-leading correction arising from a correction-to-scaling (cts) contribution,
\begin{equation}
  {\cal K}^{m_s}(T_c,H) = \kappa^{m_s} +b H^{1+(1-\beta)/\beta\delta}
  +c H^{1+(1-\beta)/\beta\delta+\omega \nu_c}
\end{equation}
we obtain 
\begin{equation}
    \kappa^{m_s} =
    \begin{cases}
    0.097(2) &,\ {\rm reg.}~ {\rm only} \\
    0.090(3) &,\ {\rm reg.} + {\rm cts}
    \end{cases}
\end{equation}
where the errors include the uncertainty on $T_c$. It is interesting to note here that the cts term is parametrically sub-leading relative to the regular term in contrast to the case for pseudo-critical temperatures in Eqs.~\ref{mixedTpc} and \ref{chiralTpc} (see Appendix \ref{app:cts} for more details).
Establishing the importance of
corrections-to-scaling terms would 
require a more detailed analysis 
of the $H$-dependence of ${\cal K}^{m_s}(T,H)$ at small values of $H$.
From both extrapolations shown in
the inset of Fig.~\ref{fig:msder_fK} we conclude
that the variation of the chiral
phase transition temperature with
the strange quark mass is rather
small. A variation of $m_s$ by 10\%
would lead to a change of $T_c$ by
less than 1\%. 

\section{Conclusions}
\label{sec:concl}

We have performed a systematic analysis of the light quark mass dependence of pseudo-critical temperatures in (2+1)-flavor QCD with the strange quark mass tuned to its physical value. 
Making use of scaling relations that describe
the dependence of the magnetic equation of state on energy-like scaling fields, {\it i.e.}
temperature and chemical potentials, close
to the chiral limit,
we determined the leading
order Taylor expansion coefficients (curvature coefficients) defining 
the dependence of the chiral phase
transition temperature on the strangeness and baryon number chemical potentials,
$T_c(\mu_B,\mu_S)=
     T_c ( 1 - (\kappa_2^B \hmu_B^2
  + \kappa_2^S \hmu_S^2 + 2\kappa_{11}^{BS} \hmu_B \hmu_S))$,
for several values of the light to strange quark mass ratio $H=m_\ell/m_s$,
and extrapolated these curvature coefficients to the chiral limit
($H=0$). We find that the curvature
coefficient decreases on lines of 
constant $\hmu_S/\hmu_B\equiv s_1$
with increasing $s_1$. On a line
corresponding to strangeness neutral matter ($s_1=0.216(3)$) it
is about 10\% smaller in than on 
a line with vanishing strangeness chemical potential ($s_1=0$).

The curvature coefficients on lines 
with fixed $s_1$ obtained on lattices with temporal extent $N_\tau=8$ 
at $T_c^{N_\tau=8}$ turn out to 
be consistent with the corresponding continuum extrapolated results obtained with
the physical light to strange quark mass
ratio, $H=1/27$, at the pseudo-critical
temperature $T_{pc}=156.5(1.5)$~MeV.
In the latter case one has
$T_{pc}^{n_S=0}(\hmu_B)= T_{pc} (1-
\kappa_2^{B,n_S=0} \hmu_B^2)$ with
$\kappa_2^{B,n_S=0}=0.012(4)$ \cite{Bazavov:2018mes}, 0.0145(25) \cite{Bonati:2018nut}, 0.0153(18)
\cite{Borsanyi:2020fev}, whereas 
in the chiral limit we obtain the not yet continuum extrapolated result
$\kappa_2^{B,n_S=0}=0.013(2)$.

All data from our calculations, presented in the figures of this
paper, can be found in Ref.~\cite{ding2024dataset}.

\textit{Note added in proof} : A paper addressing similar scenarios
in the framework of NJL model calculations appeared
shortly after our work appeared on the arXiv \cite{Ali:2024nrz}.

\section*{Acknowledgments}
This work was supported by the Deutsche Forschungsgemeinschaft
(DFG, German Research Foundation) Proj. No. 315477589-TRR 211; 
and the PUNCH4NFDI consortium
supported by the Deutsche Forschungsgemeinschaft (DFG, German Research Foundation) with project number 460248186 (PUNCH4NFDI).
It also has been supported in part by the Taiwanese NSTC project 112-2639-M-002-006-ASP;
the National Natural Science Foundation of China under Grant No. 12325508 and the National Key Research and Development Program of China under Contract No. 2022YFA1604900;
the U.S. Department of Energy, Office of Science, Office of Nuclear Physics through Contract No.~DE-SC0012704, and within the frameworks of Scientific Discovery through Advanced Computing (SciDAC) award \textit{Fundamental Nuclear Physics at the Exa\-scale and Beyond} as well as by the U.S. National Science Foundation under award PHY-2309946.

Numerical calculations have been made possible through PRACE grants
at CSCS, Switzerland.
Additional calculations have been performed on the
GPU clusters at Bielefeld University, Germany, the Central China Normal University, China, and using USQCD
resources at
the Thomas Jefferson National Accelerator Facility, USA. 

We also acknowledge very helpful discussions with Anirban Lahiri and his valuable contributions to the early stages of this work.

\appendix

\section{Finite volume effects in the vicinity of the chiral phase transition}
\label{app:finitevol}
In our previous analysis of critical behavior in
(2+1)-flavor QCD \cite{Ding:2019prx}
the region $z_{L,b} < 1$
has been found to correspond to a region where finite volume effects
are small. This can, for instance, be deduced from the $z_L$-dependence
of the finite-volume scaling functions $f_G(z,z_L), f_\chi(z,z_L)$ in the 3-d, $O(N)$
universality class. In Ref.~\cite{Karsch:2023pga} a polynomial
parametrization, valid in the vicinity of $T_c$,
{\it i.e.} for small $z$, has been given. For
$z=0$ this has a particularly simple form,
\begin{eqnarray}
    f_G(0,z_L)&=& 1+\sum_{m=4}^8 a_{0m} z_L^m
        \; , \\
        f_\chi(0,z_L)&=& \frac{1}{\delta}+\sum_{m=4}^8\left(\frac{1}{\delta}- \frac{m}{3}(1+ \frac{1}{\delta})\right) a_{0m} z_L^m
        \; ,
\end{eqnarray}
with  coefficients $a_{nm}$ given in Ref.~\cite{Karsch:2023pga}.
The scaling function $f_G(0,z_L)$ is shown in Fig.~8 of Ref.~\cite{Karsch:2023pga}. This figure
shows that finite volume effects 
at the chiral phase transition temperature, i.e., at $z=0$, rapidly increase
for $z_L\gsim 0.5-0.6$ and are negligible for $z_L \lsim 0.4$.
As $z_L=z_{0,L} z_{L,b}$, the onset of finite volume effects as
function of $z_{L,b}$, of course, depends on the non-universal
parameter $z_{0,L}$. From finite-volume fits performed for the determination
of the pseudo-critical temperature, $T_\delta$, in the vicinity of $T_c$
this non-universal scale parameter has been found to be close to 0.5.

\begin{figure}
\centering
\includegraphics[width=0.9\columnwidth]{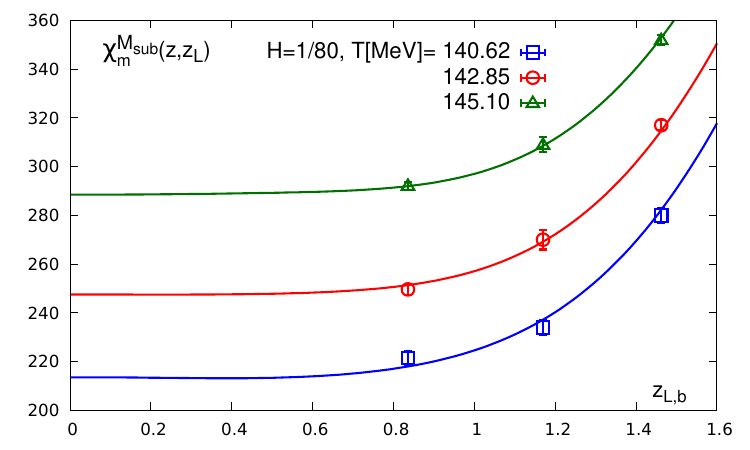}
\caption{Finite volume dependence of the chiral susceptibility obtained from the subtracted chiral condensate as defined in Eq.~\ref{chisub}. The values $z_{L,b}=0.836,1.17$ and $1.46$ correspond to lattice sizes $N_\sigma=56,40$ and $32$ respectively at fixed $N_{\tau}=8$.
Results for $N_\sigma=56$ are obtained with our
current, updated statistics.
The data points for $N_\sigma=32$ and $40$ can be found in the \href{https://journals.aps.org/prl/supplemental/10.1103/PhysRevLett.123.062002/SupplementalMaterial.pdf}{Supplementary Material} of Ref.~\cite{Ding:2019prx}.}
\label{fig:chiM_vol}
\end{figure}

We show in Fig.~\ref{fig:chiM_vol} a finite-volume fit to
$\chi_m^{M_{sub}}$ for temperatures close to the chiral
phase transition temperature $T_c$ at a fixed $H=1/80$. 
For this fit we use 
 \begin{equation}
     \chi^{M_{\rm sub}}_m =  \frac{h^{1/\delta-1}}{h_0} f_\chi(z,z_L)\; ,
      \label{chil-scaling-vol}
 \end{equation}
which is the finite volume version of the 
scaling form of the chiral susceptibility 
given in Eq.~\ref{chil-scaling}. For the scaling function
$f_\chi(z,z_L)$ we use the parametrization
given in Ref~\cite{Karsch:2023pga}.
In the fit, the chiral
phase transition temperature on lattices with temporal extent
$N_\tau=8$ has been fixed to the value $T_c=143.7$~MeV as obtained
from the scaling analysis of the chiral order parameter (see Eq.~\ref{Tcfix}).
For the remaining fit parameters $(z_0,h_0^{-1/\delta}, z_{0,L})$
we find from the fit
\begin{eqnarray}
        z_0 &=& 1.45(4)\; , \\
        h_0^{-1/\delta} &=& 39.6(2)\; , \\
        z_{0,L}&=& 0.394(3)\; ,
\end{eqnarray}
with a $\chi^2/dof=0.9$. We note that the result for
$(z_0,h_0^{-1/\delta})$ is in good agreement with the parameters obtained
from the scaling analysis of the chiral order parameter (see Eqs.~\ref{z0fix},\ref{h0fix}).
From this we deduce that in (2+1)-flavor QCD studies close to $T_c$
finite volume effects are small for $z_{L,b}\simeq 1$ in observables of interest 
in our current analysis.

\section{Corrections-to-Scaling and regular contributions}
\label{app:cts}

The leading singular contribution to the free energy density has been introduced in Eq.~\ref{fs}. Sub-dominant, still
divergent, universal corrections to this arise from from another scaling field, $u_3,$ in the singular part of the free energy density,
\begin{equation}
 f_s(z)\Rightarrow  f_s(z,u_3) \equiv h_0 h^{1+1/\delta} f_f( z,u_3 h^{\omega \nu_c})\; ,
\end{equation}
with $\nu_c\equiv \nu/\beta\delta$,
with $\omega=0.79$, $\nu=0.6723$.

This can be used as a starting point for a scaling analysis that
takes into account corrections to scaling. The magnetization then
is given by
\begin{eqnarray}
M_\ell= -h^{1/\delta} &&\Bigl( (1+1/\delta) f_f(z, u_3 h^{\omega \nu_c})
\nonumber \\
&&- \frac{1}{\beta\delta}  z f^{(1,0)}_f(z,u_3 h^{\omega \nu_c})
\nonumber \\
&& + u_3 \omega\nu_c h^{\omega \nu_c} f^{(0,1)}_f (z,u_3 h^{\omega \nu_c})
\Bigr) \; .
\end{eqnarray}
Here $f^{(1,0)}_f$ and $f^{(0,1)}_f$ denote the derivatives of $f_f$ with respect
to the first and second argument,
respectively.

For small $h$, {\it i.e.} close to the chiral limit, one may expand $f_f$ in terms of $u_3$.
Keeping only the leading order correction term one gets,
\begin{eqnarray}
M_\ell h^{-1/\delta} &=& f_G(z) - u_3 h^{\omega \nu_c} f_{G,cts}(z) +
{\cal O}(h^{2 \omega \nu_c}) \; ,
\label{Mb}
\end{eqnarray}
where $f_G(z)$ has been introduced in Eq.~\ref{fG} and 
$f_{G,cts}(z)\equiv (1+1/\delta +\omega \nu_c ) f_f^{(0,1)}(z,0)-\frac{z}{\beta \delta} f_f^{(1,1)}(z,0)$.

In addition to the correction-to-scaling (cts) term we 
also consider a 
contribution from regular terms, 
\begin{eqnarray}
M_\ell  &=& h^{1/\delta}\left( f_G(z) - u_3 h^{\omega \nu_c} f_{G,cts}(z) 
\right) 
\nonumber \\
&&+ H (a_0+a_1 (T-T_c)+ \frac{a_2}{2} (T-T_c)^2 +...)\; .
\label{Mbs}
\end{eqnarray}
 
From this one obtains for the 
mixed susceptibility $\chi^{M_\ell}_{t(T)}$,
\begin{eqnarray}
\chi^{M_\ell}_{t(T)}&=& -z_0 h_0^{-1/\delta} H^{(\beta-1)/\beta\delta} \Bigl( f'_G(z) 
\nonumber \\
&&- u_3 h_0^{-\omega \nu_c} H^{\omega \nu_c} f'_{G,cts}(z) \Bigr) 
\nonumber \\
&&+H ( a_1 +a_2 (T-T_c)+...) \; .
\label{chits}
\end{eqnarray}
In the absence of contributions
from regular and cts terms, $\chi^{M_\ell}_{t(T)}$
has a maximum at a pseudo-critical temperature
$T_{pc,t}$ given by the condition $z=z_t$. Corrections to this result will
be small for small $H$. We therefore expand the right hand side of 
Eq.~\ref{chits} around the maximum
of $f'_G(z)$, i.e. around $z_t$,
\begin{eqnarray}
\chi^{M_\ell}_{t(T)} &=& -z_0 h_0^{-1/\delta} H^{(\beta-1)/\beta\delta} \Bigl(  
f'_G(z_t) +\frac{1}{2}f'''_G(z_t) (z-z_t)^2  \nonumber \\
&& - u_3 h_0^{-\omega \nu_c} H^{\omega \nu_c} (f'_{G,cts}(z_t)+f''_{G,cts}(z_t)(z-z_t)) \Bigr) 
\nonumber \\
&&+H (a_1 + a_2 (T-T_c) ) \; .
\end{eqnarray}
With this we determine the location
of the maximum of $\chi^{M_\ell}_{t(T)}(T,H)$,
\begin{eqnarray}
0=\frac{\partial \chi^{M_\ell}_{t(T)}(T,H)}{\partial T}&=&
-\frac{z_0^2 h_0^{-1/\delta}}{T_c}  H^{(\beta-2)/\beta\delta} \Bigl(  
f'''_G(z_t) (z-z_t)  \nonumber \\
&& - u_3 h_0^{-\omega \nu_c} H^{\omega \nu_c} f''_{G,cts}(z_t) \Bigr) 
 +H a_2 \;
\nonumber \\
\Rightarrow  T_{pc,t}(H)=T_c\Bigl(&1& + \frac{z_t}{z_0} H^{1/\beta\delta} + B_t H^{(1+\omega \nu)/\beta\delta} 
\nonumber \\
&+& C_t a_2 H^{1+(3-\beta)/\beta\delta}
    \Bigr) \; .
\end{eqnarray}
Similarly one finds for the 
pseudo-critical temperature $T_{pc,m}$, determined from 
the maximum of $\chi^{M_{\rm sub}}_m$, 
\cite{Bazavov:2011nk}
\begin{eqnarray}
    T_{pc,m}(H)=T_c\Bigl( 1  &+& \frac{z_m}{z_0}  H^{1/\beta\delta} + B_m H^{(1+\omega \nu)/\beta\delta}
    \nonumber \\
    &+&  C_m a_1  H^{1+(2-\beta)/\beta\delta}
    \Bigr)
\end{eqnarray}
Note that regular contributions are 
suppressed relative to the corrections-to-scaling terms in
$T_{pc,t}$ and $T_{pc,m}$.

The curvature coefficients are obtained from ratios of susceptibilities, e.g. like ${\cal K}_2^f(T,H)$ given in  Eq.~\ref{kapparatio}. 
These ratios will be evaluated at or
close to $T_c$.
In the presence 
of corrections-to-scaling the scaling function $f'_G(z)$, appearing in that
equation, gets replaced by
$f'_G(z) - u_3 h^{\omega \nu_c} f'_{G,cts}(z)$. The cts term thus
multiplies the dominant divergent term and is proportional to the 
curvature coefficient, e.g. $\kappa_2^f$. 

At $T_c$ one obtains  
\begin{widetext}
\begin{eqnarray}
{\cal K}_2^f(T_c,H)
&=& \frac{\kappa_2^f h^{(\beta-1)/\beta\delta}(f'_G(0) - \tilde{u}_3 H^{\omega \nu_c} f'_{G,cts}(0)) /t_0 + H a_f}{h^{(\beta-1)/\beta\delta}(f'_G(0) - \tilde{u}_3 H^{\omega \nu_c} f'_{G,cts}(0)) /t_0 + H a_t } \; ,
        \label{kapparatio2} 
\end{eqnarray}
%\end{widetext}
with $\tilde{u}_3=h_0^{-\omega\nu_c} u_3$.
In the absence of a regular term the
cts contribution thus cancels and 
there is no $H$-dependent correction
to $\kappa_2^f$. As a consequence the cts corrections becomes sub-leading
to the regular term, i.e.
%\begin{widetext}
\begin{eqnarray}
{\cal K}_2^f(T_c,H)
    &=& \frac{\kappa_2^f + \tilde{a}_f  H^{1+(1-\beta)/\beta\delta}/(f'_G(0) - \tilde{u}_3 H^{\omega \nu_c} f'_{G,cts}(0))}{1+\tilde{a}_t H^{1+(1-\beta)/\beta\delta}/(f'_G(0) - \tilde{u}_3 H^{\omega \nu_c} f'_{G,cts}(0))}
        \label{kapparatio3}  
        \nonumber \\
&=& \kappa_2^f +A_f H^{1+(1-\beta)/\beta\delta}+ B_f
H^{1+(1-\beta)/\beta\delta +\omega \nu_c} \; .
        \label{kapparatio4} 
\end{eqnarray}
\end{widetext}

\section{Strangeness neutrality in the chiral limit}
\label{app:s1}
We give here results for the quark mass dependence of the
leading order strangeness neutrality relation between the strangeness and baryon number
chemical potentials.
\begin{figure}[t]
\centering
\includegraphics[width=0.48\textwidth]{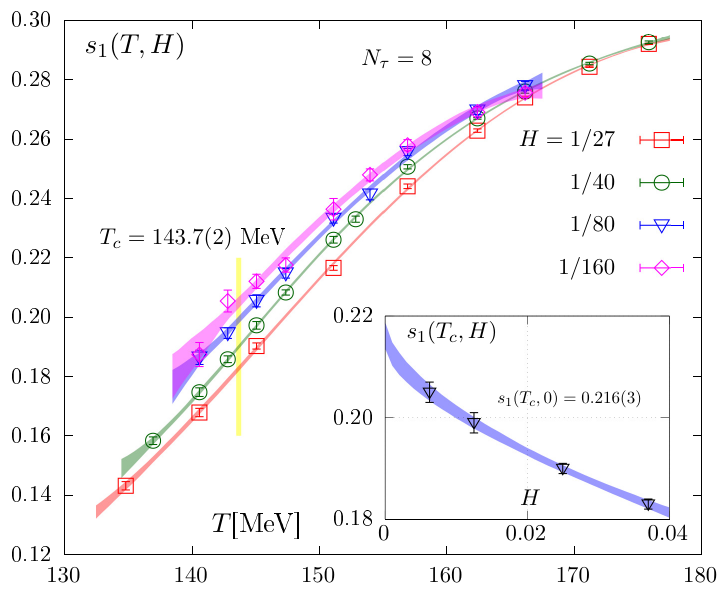}
\caption{Light quark mass dependence of the leading order strangeness neutrality coefficient $s_1(T,H)$ versus temperature.
The inset shows $s_1(T_c,H)$ versus $H$. Here the band gives a 
chiral extrapolation using the scaling ansatz given in Eq.~\ref{scale}.
The yellow band shows $T_c$ as given in Eq.~\ref{Tcfix}.
}
\label{fig:s1}
\end{figure}

Expanding the net strangeness number density, $n_S$, to leading order in the chemical potentials $(\hmu_B, \hmu_Q, \hmu_S)$ gives,
\begin{equation}
    n_S= \chi_{11}^{BS}\hmu_B+\chi_{11}^{QS}\hmu_Q+\chi_{2}^{S}\hmu_S \; ,
\label{s1}
\end{equation}
where $\chi_{11}^{XS}$, with $X=B, Q$, and
$\chi_2^S$ are second order conserved charge cumulants. Their definition and recent continuum extrapolated results obtained in calculations with the HISQ action are given in \cite{Bollweg:2021vqf}. 
In the case of vanishing electric charge chemical potential, $\hmu_Q=0$, which
is considered throughout this work, one thus finds from Eq.~\ref{s1} that
strangeness and baryon number chemical potentials are related through,

\begin{equation}
    \hmu_S= s_1(T,H)\hmu_B\; ,
\end{equation}
with 
\begin{equation}
    s_1(T,H)= - \frac{\chi_{11}^{BS}}{\chi_2^S}\; .
\end{equation}
Continuum extrapolated results for this ratio at physical values of the light and strange quark masses, {\it i.e.} for $H=1/27$ have also been reported in \cite{Bollweg:2021vqf}
\begin{eqnarray}
    s_1 (T_{pc},1/27) &=&
0.236(5)(6) \; .
\end{eqnarray}
Here $T_{pc}$ is obtained by averaging 
over continuum extrapolated results for $T_{pc,m}$ and $T_{pc,t}$ \cite{Bazavov:2018mes}.
For the analysis presented 
in this work we are interested in the chiral limit results at the chiral 
phase transition temperature on lattices with fixed $N_\tau=8$, {\it i.e.}, we
want to determine $s_1(T_c,0)$.
In Fig.~\ref{fig:s1} we show
results for $s_1(T,H)$ obtained on lattices with
temporal extent $N_\tau=8$
as function of temperature 
and quark mass ratio $H$.
In the inset of Fig.~\ref{fig:s1} we show $s_1(T_c,H)$. A chiral extrapolation of these data, using
the scaling ansatz appropriate for
the scaling of  energy-like observables in the $3d$, $O(N)$ universality classes \cite{Clarke:2020htu},
\begin{equation}
    s_1(T_c,H) = s_1(T_c,0) + c H^{1+(\beta-1)/\beta\delta} \; ,
    \label{scale}
\end{equation}
yields 
\begin{equation}
    s_1(T_c,0) = 0.216(3) \; ,
\end{equation}
where the error includes the statistical error and the uncertainty arising from the error on $T_c$. 
In order to provide further support 
for the use of a scaling ansatz 
for the chiral extrapolation, clearly
more data at different values of $H$
are needed.

%\bibliography{chiral_curvature}

\begin{thebibliography}{49}%
\makeatletter
\providecommand \@ifxundefined [1]{%
 \@ifx{#1\undefined}
}%
\providecommand \@ifnum [1]{%
 \ifnum #1\expandafter \@firstoftwo
 \else \expandafter \@secondoftwo
 \fi
}%
\providecommand \@ifx [1]{%
 \ifx #1\expandafter \@firstoftwo
 \else \expandafter \@secondoftwo
 \fi
}%
\providecommand \natexlab [1]{#1}%
\providecommand \enquote  [1]{``#1''}%
\providecommand \bibnamefont  [1]{#1}%
\providecommand \bibfnamefont [1]{#1}%
\providecommand \citenamefont [1]{#1}%
\providecommand \href@noop [0]{\@secondoftwo}%
\providecommand \href [0]{\begingroup \@sanitize@url \@href}%
\providecommand \@href[1]{\@@startlink{#1}\@@href}%
\providecommand \@@href[1]{\endgroup#1\@@endlink}%
\providecommand \@sanitize@url [0]{\catcode `\\12\catcode `\$12\catcode
  `\&12\catcode `\#12\catcode `\^12\catcode `\_12\catcode `\%12\relax}%
\providecommand \@@startlink[1]{}%
\providecommand \@@endlink[0]{}%
\providecommand \url  [0]{\begingroup\@sanitize@url \@url }%
\providecommand \@url [1]{\endgroup\@href {#1}{\urlprefix }}%
\providecommand \urlprefix  [0]{URL }%
\providecommand \Eprint [0]{\href }%
\providecommand \doibase [0]{http://dx.doi.org/}%
\providecommand \selectlanguage [0]{\@gobble}%
\providecommand \bibinfo  [0]{\@secondoftwo}%
\providecommand \bibfield  [0]{\@secondoftwo}%
\providecommand \translation [1]{[#1]}%
\providecommand \BibitemOpen [0]{}%
\providecommand \bibitemStop [0]{}%
\providecommand \bibitemNoStop [0]{.\EOS\space}%
\providecommand \EOS [0]{\spacefactor3000\relax}%
\providecommand \BibitemShut  [1]{\csname bibitem#1\endcsname}%
\let\auto@bib@innerbib\@empty
%</preamble>
\bibitem [{\citenamefont {Bazavov}\ \emph {et~al.}(2019)\citenamefont {Bazavov}
  \emph {et~al.}}]{Bazavov:2018mes}%
  \BibitemOpen
  \bibfield  {author} {\bibinfo {author} {\bibfnamefont {A.}~\bibnamefont
  {Bazavov}} \emph {et~al.} (\bibinfo {collaboration} {HotQCD}),\ }\href
  {\doibase 10.1016/j.physletb.2019.05.013} {\bibfield  {journal} {\bibinfo
  {journal} {Phys.\ Lett.\ B}\ }\textbf {\bibinfo {volume} {795}},\ \bibinfo
  {pages} {15} (\bibinfo {year} {2019})},\ \Eprint
  {http://arxiv.org/abs/1812.08235} {arXiv:1812.08235 [hep-lat]} \BibitemShut
  {NoStop}%
\bibitem [{\citenamefont {Bhattacharya}\ \emph {et~al.}(2014)\citenamefont
  {Bhattacharya} \emph {et~al.}}]{Bhattacharya:2014ara}%
  \BibitemOpen
  \bibfield  {author} {\bibinfo {author} {\bibfnamefont {T.}~\bibnamefont
  {Bhattacharya}} \emph {et~al.},\ }\href {\doibase
  10.1103/PhysRevLett.113.082001} {\bibfield  {journal} {\bibinfo  {journal}
  {Phys. Rev. Lett.}\ }\textbf {\bibinfo {volume} {113}},\ \bibinfo {pages}
  {082001} (\bibinfo {year} {2014})},\ \Eprint {http://arxiv.org/abs/1402.5175}
  {arXiv:1402.5175 [hep-lat]} \BibitemShut {NoStop}%
\bibitem [{\citenamefont {Borsanyi}\ \emph {et~al.}(2020)\citenamefont
  {Borsanyi}, \citenamefont {Fodor}, \citenamefont {Guenther}, \citenamefont
  {Kara}, \citenamefont {Katz}, \citenamefont {Parotto}, \citenamefont
  {Pasztor}, \citenamefont {Ratti},\ and\ \citenamefont
  {Szabo}}]{Borsanyi:2020fev}%
  \BibitemOpen
  \bibfield  {author} {\bibinfo {author} {\bibfnamefont {S.}~\bibnamefont
  {Borsanyi}}, \bibinfo {author} {\bibfnamefont {Z.}~\bibnamefont {Fodor}},
  \bibinfo {author} {\bibfnamefont {J.~N.}\ \bibnamefont {Guenther}}, \bibinfo
  {author} {\bibfnamefont {R.}~\bibnamefont {Kara}}, \bibinfo {author}
  {\bibfnamefont {S.~D.}\ \bibnamefont {Katz}}, \bibinfo {author}
  {\bibfnamefont {P.}~\bibnamefont {Parotto}}, \bibinfo {author} {\bibfnamefont
  {A.}~\bibnamefont {Pasztor}}, \bibinfo {author} {\bibfnamefont
  {C.}~\bibnamefont {Ratti}}, \ and\ \bibinfo {author} {\bibfnamefont {K.~K.}\
  \bibnamefont {Szabo}},\ }\href {\doibase 10.1103/PhysRevLett.125.052001}
  {\bibfield  {journal} {\bibinfo  {journal} {Phys. Rev. Lett.}\ }\textbf
  {\bibinfo {volume} {125}},\ \bibinfo {pages} {052001} (\bibinfo {year}
  {2020})},\ \Eprint {http://arxiv.org/abs/2002.02821} {arXiv:2002.02821
  [hep-lat]} \BibitemShut {NoStop}%
\bibitem [{\citenamefont {Ding}\ \emph {et~al.}(2019)\citenamefont {Ding} \emph
  {et~al.}}]{Ding:2019prx}%
  \BibitemOpen
  \bibfield  {author} {\bibinfo {author} {\bibfnamefont {H.}~\bibnamefont
  {Ding}} \emph {et~al.},\ }\href {\doibase 10.1103/PhysRevLett.123.062002}
  {\bibfield  {journal} {\bibinfo  {journal} {Phys. Rev. Lett.}\ }\textbf
  {\bibinfo {volume} {123}},\ \bibinfo {pages} {062002} (\bibinfo {year}
  {2019})},\ \Eprint {http://arxiv.org/abs/1903.04801} {arXiv:1903.04801
  [hep-lat]} \BibitemShut {NoStop}%
\bibitem [{\citenamefont {Kotov}\ \emph {et~al.}(2021)\citenamefont {Kotov},
  \citenamefont {Lombardo},\ and\ \citenamefont {Trunin}}]{Kotov:2021rah}%
  \BibitemOpen
  \bibfield  {author} {\bibinfo {author} {\bibfnamefont {A.~Y.}\ \bibnamefont
  {Kotov}}, \bibinfo {author} {\bibfnamefont {M.~P.}\ \bibnamefont {Lombardo}},
  \ and\ \bibinfo {author} {\bibfnamefont {A.}~\bibnamefont {Trunin}},\ }\href
  {\doibase 10.1016/j.physletb.2021.136749} {\bibfield  {journal} {\bibinfo
  {journal} {Phys. Lett. B}\ }\textbf {\bibinfo {volume} {823}},\ \bibinfo
  {pages} {136749} (\bibinfo {year} {2021})},\ \Eprint
  {http://arxiv.org/abs/2105.09842} {arXiv:2105.09842 [hep-lat]} \BibitemShut
  {NoStop}%
\bibitem [{\citenamefont {Braun}\ \emph {et~al.}(2023)\citenamefont {Braun}
  \emph {et~al.}}]{Braun:2023qak}%
  \BibitemOpen
  \bibfield  {author} {\bibinfo {author} {\bibfnamefont {J.}~\bibnamefont
  {Braun}} \emph {et~al.},\ }\href@noop {} {\  (\bibinfo {year} {2023})},\
  \Eprint {http://arxiv.org/abs/2310.19853} {arXiv:2310.19853 [hep-ph]}
  \BibitemShut {NoStop}%
\bibitem [{\citenamefont {Cuteri}\ \emph {et~al.}(2021)\citenamefont {Cuteri},
  \citenamefont {Philipsen},\ and\ \citenamefont {Sciarra}}]{Cuteri:2021ikv}%
  \BibitemOpen
  \bibfield  {author} {\bibinfo {author} {\bibfnamefont {F.}~\bibnamefont
  {Cuteri}}, \bibinfo {author} {\bibfnamefont {O.}~\bibnamefont {Philipsen}}, \
  and\ \bibinfo {author} {\bibfnamefont {A.}~\bibnamefont {Sciarra}},\ }\href
  {\doibase 10.1007/JHEP11(2021)141} {\bibfield  {journal} {\bibinfo  {journal}
  {JHEP}\ }\textbf {\bibinfo {volume} {11}},\ \bibinfo {pages} {141} (\bibinfo
  {year} {2021})},\ \Eprint {http://arxiv.org/abs/2107.12739} {arXiv:2107.12739
  [hep-lat]} \BibitemShut {NoStop}%
\bibitem [{\citenamefont {Halasz}\ \emph {et~al.}(1998)\citenamefont {Halasz},
  \citenamefont {Jackson}, \citenamefont {Shrock}, \citenamefont {Stephanov},\
  and\ \citenamefont {Verbaarschot}}]{Halasz:1998qr}%
  \BibitemOpen
  \bibfield  {author} {\bibinfo {author} {\bibfnamefont {A.~M.}\ \bibnamefont
  {Halasz}}, \bibinfo {author} {\bibfnamefont {A.~D.}\ \bibnamefont {Jackson}},
  \bibinfo {author} {\bibfnamefont {R.~E.}\ \bibnamefont {Shrock}}, \bibinfo
  {author} {\bibfnamefont {M.~A.}\ \bibnamefont {Stephanov}}, \ and\ \bibinfo
  {author} {\bibfnamefont {J.~J.~M.}\ \bibnamefont {Verbaarschot}},\ }\href
  {\doibase 10.1103/PhysRevD.58.096007} {\bibfield  {journal} {\bibinfo
  {journal} {Phys. Rev. D}\ }\textbf {\bibinfo {volume} {58}},\ \bibinfo
  {pages} {096007} (\bibinfo {year} {1998})},\ \Eprint
  {http://arxiv.org/abs/hep-ph/9804290} {arXiv:hep-ph/9804290} \BibitemShut
  {NoStop}%
\bibitem [{\citenamefont {Hatta}\ and\ \citenamefont
  {Ikeda}(2003)}]{Hatta:2002sj}%
  \BibitemOpen
  \bibfield  {author} {\bibinfo {author} {\bibfnamefont {Y.}~\bibnamefont
  {Hatta}}\ and\ \bibinfo {author} {\bibfnamefont {T.}~\bibnamefont {Ikeda}},\
  }\href {\doibase 10.1103/PhysRevD.67.014028} {\bibfield  {journal} {\bibinfo
  {journal} {Phys. Rev. D}\ }\textbf {\bibinfo {volume} {67}},\ \bibinfo
  {pages} {014028} (\bibinfo {year} {2003})},\ \Eprint
  {http://arxiv.org/abs/hep-ph/0210284} {arXiv:hep-ph/0210284} \BibitemShut
  {NoStop}%
\bibitem [{\citenamefont {Stephanov}(2006)}]{Stephanov:2006dn}%
  \BibitemOpen
  \bibfield  {author} {\bibinfo {author} {\bibfnamefont {M.~A.}\ \bibnamefont
  {Stephanov}},\ }\href {\doibase 10.1103/PhysRevD.73.094508} {\bibfield
  {journal} {\bibinfo  {journal} {Phys. Rev. D}\ }\textbf {\bibinfo {volume}
  {73}},\ \bibinfo {pages} {094508} (\bibinfo {year} {2006})},\ \Eprint
  {http://arxiv.org/abs/hep-lat/0603014} {arXiv:hep-lat/0603014} \BibitemShut
  {NoStop}%
\bibitem [{\citenamefont {Buballa}\ and\ \citenamefont
  {Carignano}(2019)}]{Buballa:2018hux}%
  \BibitemOpen
  \bibfield  {author} {\bibinfo {author} {\bibfnamefont {M.}~\bibnamefont
  {Buballa}}\ and\ \bibinfo {author} {\bibfnamefont {S.}~\bibnamefont
  {Carignano}},\ }\href {\doibase 10.1016/j.physletb.2019.02.045} {\bibfield
  {journal} {\bibinfo  {journal} {Phys. Lett. B}\ }\textbf {\bibinfo {volume}
  {791}},\ \bibinfo {pages} {361} (\bibinfo {year} {2019})},\ \Eprint
  {http://arxiv.org/abs/1809.10066} {arXiv:1809.10066 [hep-ph]} \BibitemShut
  {NoStop}%
\bibitem [{\citenamefont {Pisarski}\ and\ \citenamefont
  {Wilczek}(1984)}]{Pisarski:1983ms}%
  \BibitemOpen
  \bibfield  {author} {\bibinfo {author} {\bibfnamefont {R.~D.}\ \bibnamefont
  {Pisarski}}\ and\ \bibinfo {author} {\bibfnamefont {F.}~\bibnamefont
  {Wilczek}},\ }\href {\doibase 10.1103/PhysRevD.29.338} {\bibfield  {journal}
  {\bibinfo  {journal} {Phys.\ Rev.\ D}\ }\textbf {\bibinfo {volume} {29}},\
  \bibinfo {pages} {338} (\bibinfo {year} {1984})}\BibitemShut {NoStop}%
\bibitem [{\citenamefont {Butti}\ \emph {et~al.}(2003)\citenamefont {Butti},
  \citenamefont {Pelissetto},\ and\ \citenamefont {Vicari}}]{Butti:2003nu}%
  \BibitemOpen
  \bibfield  {author} {\bibinfo {author} {\bibfnamefont {A.}~\bibnamefont
  {Butti}}, \bibinfo {author} {\bibfnamefont {A.}~\bibnamefont {Pelissetto}}, \
  and\ \bibinfo {author} {\bibfnamefont {E.}~\bibnamefont {Vicari}},\ }\href
  {\doibase 10.1088/1126-6708/2003/08/029} {\bibfield  {journal} {\bibinfo
  {journal} {JHEP}\ }\textbf {\bibinfo {volume} {08}},\ \bibinfo {pages} {029}
  (\bibinfo {year} {2003})},\ \Eprint {http://arxiv.org/abs/hep-ph/0307036}
  {arXiv:hep-ph/0307036} \BibitemShut {NoStop}%
\bibitem [{\citenamefont {Buchoff}\ \emph {et~al.}(2014)\citenamefont {Buchoff}
  \emph {et~al.}}]{Buchoff:2013nra}%
  \BibitemOpen
  \bibfield  {author} {\bibinfo {author} {\bibfnamefont {M.~I.}\ \bibnamefont
  {Buchoff}} \emph {et~al.},\ }\href {\doibase 10.1103/PhysRevD.89.054514}
  {\bibfield  {journal} {\bibinfo  {journal} {Phys. Rev. D}\ }\textbf {\bibinfo
  {volume} {89}},\ \bibinfo {pages} {054514} (\bibinfo {year} {2014})},\
  \Eprint {http://arxiv.org/abs/1309.4149} {arXiv:1309.4149 [hep-lat]}
  \BibitemShut {NoStop}%
\bibitem [{\citenamefont {Pelissetto}\ and\ \citenamefont
  {Vicari}(2013)}]{Pelissetto:2013hqa}%
  \BibitemOpen
  \bibfield  {author} {\bibinfo {author} {\bibfnamefont {A.}~\bibnamefont
  {Pelissetto}}\ and\ \bibinfo {author} {\bibfnamefont {E.}~\bibnamefont
  {Vicari}},\ }\href {\doibase 10.1103/PhysRevD.88.105018} {\bibfield
  {journal} {\bibinfo  {journal} {Phys. Rev. D}\ }\textbf {\bibinfo {volume}
  {88}},\ \bibinfo {pages} {105018} (\bibinfo {year} {2013})},\ \Eprint
  {http://arxiv.org/abs/1309.5446} {arXiv:1309.5446 [hep-lat]} \BibitemShut
  {NoStop}%
\bibitem [{\citenamefont {Resch}\ \emph {et~al.}(2019)\citenamefont {Resch},
  \citenamefont {Rennecke},\ and\ \citenamefont {Schaefer}}]{Resch:2017vjs}%
  \BibitemOpen
  \bibfield  {author} {\bibinfo {author} {\bibfnamefont {S.}~\bibnamefont
  {Resch}}, \bibinfo {author} {\bibfnamefont {F.}~\bibnamefont {Rennecke}}, \
  and\ \bibinfo {author} {\bibfnamefont {B.-J.}\ \bibnamefont {Schaefer}},\
  }\href {\doibase 10.1103/PhysRevD.99.076005} {\bibfield  {journal} {\bibinfo
  {journal} {Phys. Rev. D}\ }\textbf {\bibinfo {volume} {99}},\ \bibinfo
  {pages} {076005} (\bibinfo {year} {2019})},\ \Eprint
  {http://arxiv.org/abs/1712.07961} {arXiv:1712.07961 [hep-ph]} \BibitemShut
  {NoStop}%
\bibitem [{\citenamefont {Aoki}\ \emph {et~al.}(2021)\citenamefont {Aoki},
  \citenamefont {Aoki}, \citenamefont {Cossu}, \citenamefont {Fukaya},
  \citenamefont {Hashimoto}, \citenamefont {Kaneko}, \citenamefont
  {Rohrhofer},\ and\ \citenamefont {Suzuki}}]{Aoki:2020noz}%
  \BibitemOpen
  \bibfield  {author} {\bibinfo {author} {\bibfnamefont {S.}~\bibnamefont
  {Aoki}}, \bibinfo {author} {\bibfnamefont {Y.}~\bibnamefont {Aoki}}, \bibinfo
  {author} {\bibfnamefont {G.}~\bibnamefont {Cossu}}, \bibinfo {author}
  {\bibfnamefont {H.}~\bibnamefont {Fukaya}}, \bibinfo {author} {\bibfnamefont
  {S.}~\bibnamefont {Hashimoto}}, \bibinfo {author} {\bibfnamefont
  {T.}~\bibnamefont {Kaneko}}, \bibinfo {author} {\bibfnamefont
  {C.}~\bibnamefont {Rohrhofer}}, \ and\ \bibinfo {author} {\bibfnamefont
  {K.}~\bibnamefont {Suzuki}} (\bibinfo {collaboration} {JLQCD}),\ }\href
  {\doibase 10.1103/PhysRevD.103.074506} {\bibfield  {journal} {\bibinfo
  {journal} {Phys. Rev. D}\ }\textbf {\bibinfo {volume} {103}},\ \bibinfo
  {pages} {074506} (\bibinfo {year} {2021})},\ \Eprint
  {http://arxiv.org/abs/2011.01499} {arXiv:2011.01499 [hep-lat]} \BibitemShut
  {NoStop}%
\bibitem [{\citenamefont {Ding}\ \emph {et~al.}(2021)\citenamefont {Ding},
  \citenamefont {Li}, \citenamefont {Mukherjee}, \citenamefont {Tomiya},
  \citenamefont {Wang},\ and\ \citenamefont {Zhang}}]{Ding:2020xlj}%
  \BibitemOpen
  \bibfield  {author} {\bibinfo {author} {\bibfnamefont {H.~T.}\ \bibnamefont
  {Ding}}, \bibinfo {author} {\bibfnamefont {S.~T.}\ \bibnamefont {Li}},
  \bibinfo {author} {\bibfnamefont {S.}~\bibnamefont {Mukherjee}}, \bibinfo
  {author} {\bibfnamefont {A.}~\bibnamefont {Tomiya}}, \bibinfo {author}
  {\bibfnamefont {X.~D.}\ \bibnamefont {Wang}}, \ and\ \bibinfo {author}
  {\bibfnamefont {Y.}~\bibnamefont {Zhang}},\ }\href {\doibase
  10.1103/PhysRevLett.126.082001} {\bibfield  {journal} {\bibinfo  {journal}
  {Phys. Rev. Lett.}\ }\textbf {\bibinfo {volume} {126}},\ \bibinfo {pages}
  {082001} (\bibinfo {year} {2021})},\ \Eprint
  {http://arxiv.org/abs/2010.14836} {arXiv:2010.14836 [hep-lat]} \BibitemShut
  {NoStop}%
\bibitem [{\citenamefont {Ding}\ \emph {et~al.}(2023)\citenamefont {Ding},
  \citenamefont {Huang}, \citenamefont {Mukherjee},\ and\ \citenamefont
  {Petreczky}}]{Ding:2023oxy}%
  \BibitemOpen
  \bibfield  {author} {\bibinfo {author} {\bibfnamefont {H.-T.}\ \bibnamefont
  {Ding}}, \bibinfo {author} {\bibfnamefont {W.-P.}\ \bibnamefont {Huang}},
  \bibinfo {author} {\bibfnamefont {S.}~\bibnamefont {Mukherjee}}, \ and\
  \bibinfo {author} {\bibfnamefont {P.}~\bibnamefont {Petreczky}},\ }\href
  {\doibase 10.1103/PhysRevLett.131.161903} {\bibfield  {journal} {\bibinfo
  {journal} {Phys. Rev. Lett.}\ }\textbf {\bibinfo {volume} {131}},\ \bibinfo
  {pages} {161903} (\bibinfo {year} {2023})},\ \Eprint
  {http://arxiv.org/abs/2305.10916} {arXiv:2305.10916 [hep-lat]} \BibitemShut
  {NoStop}%
\bibitem [{\citenamefont {Kovacs}(2023)}]{Kovacs:2023vzi}%
  \BibitemOpen
  \bibfield  {author} {\bibinfo {author} {\bibfnamefont {T.~G.}\ \bibnamefont
  {Kovacs}},\ }\href@noop {} {\  (\bibinfo {year} {2023})},\ \Eprint
  {http://arxiv.org/abs/2311.04208} {arXiv:2311.04208 [hep-lat]} \BibitemShut
  {NoStop}%
\bibitem [{\citenamefont {Karsch}(2019)}]{Karsch:2019mbv}%
  \BibitemOpen
  \bibfield  {author} {\bibinfo {author} {\bibfnamefont {F.}~\bibnamefont
  {Karsch}},\ }\href {\doibase 10.22323/1.347.0163} {\bibfield  {journal}
  {\bibinfo  {journal} {PoS}\ }\textbf {\bibinfo {volume} {CORFU2018}},\
  \bibinfo {pages} {163} (\bibinfo {year} {2019})},\ \Eprint
  {http://arxiv.org/abs/1905.03936} {arXiv:1905.03936 [hep-lat]} \BibitemShut
  {NoStop}%
\bibitem [{\citenamefont {Allton}\ \emph {et~al.}(2002)\citenamefont {Allton},
  \citenamefont {Ejiri}, \citenamefont {Hands}, \citenamefont {Kaczmarek},
  \citenamefont {Karsch}, \citenamefont {Laermann}, \citenamefont {Schmidt},\
  and\ \citenamefont {Scorzato}}]{Allton:2002zi}%
  \BibitemOpen
  \bibfield  {author} {\bibinfo {author} {\bibfnamefont {C.~R.}\ \bibnamefont
  {Allton}}, \bibinfo {author} {\bibfnamefont {S.}~\bibnamefont {Ejiri}},
  \bibinfo {author} {\bibfnamefont {S.~J.}\ \bibnamefont {Hands}}, \bibinfo
  {author} {\bibfnamefont {O.}~\bibnamefont {Kaczmarek}}, \bibinfo {author}
  {\bibfnamefont {F.}~\bibnamefont {Karsch}}, \bibinfo {author} {\bibfnamefont
  {E.}~\bibnamefont {Laermann}}, \bibinfo {author} {\bibfnamefont
  {C.}~\bibnamefont {Schmidt}}, \ and\ \bibinfo {author} {\bibfnamefont
  {L.}~\bibnamefont {Scorzato}},\ }\href {\doibase 10.1103/PhysRevD.66.074507}
  {\bibfield  {journal} {\bibinfo  {journal} {Phys. Rev. D}\ }\textbf {\bibinfo
  {volume} {66}},\ \bibinfo {pages} {074507} (\bibinfo {year} {2002})},\
  \Eprint {http://arxiv.org/abs/hep-lat/0204010} {arXiv:hep-lat/0204010}
  \BibitemShut {NoStop}%
\bibitem [{\citenamefont {Kaczmarek}\ \emph {et~al.}(2011)\citenamefont
  {Kaczmarek}, \citenamefont {Karsch}, \citenamefont {Laermann}, \citenamefont
  {Miao}, \citenamefont {Mukherjee}, \citenamefont {Petreczky}, \citenamefont
  {Schmidt}, \citenamefont {Soeldner},\ and\ \citenamefont
  {Unger}}]{Kaczmarek:2011zz}%
  \BibitemOpen
  \bibfield  {author} {\bibinfo {author} {\bibfnamefont {O.}~\bibnamefont
  {Kaczmarek}}, \bibinfo {author} {\bibfnamefont {F.}~\bibnamefont {Karsch}},
  \bibinfo {author} {\bibfnamefont {E.}~\bibnamefont {Laermann}}, \bibinfo
  {author} {\bibfnamefont {C.}~\bibnamefont {Miao}}, \bibinfo {author}
  {\bibfnamefont {S.}~\bibnamefont {Mukherjee}}, \bibinfo {author}
  {\bibfnamefont {P.}~\bibnamefont {Petreczky}}, \bibinfo {author}
  {\bibfnamefont {C.}~\bibnamefont {Schmidt}}, \bibinfo {author} {\bibfnamefont
  {W.}~\bibnamefont {Soeldner}}, \ and\ \bibinfo {author} {\bibfnamefont
  {W.}~\bibnamefont {Unger}},\ }\href {\doibase 10.1103/PhysRevD.83.014504}
  {\bibfield  {journal} {\bibinfo  {journal} {Phys. Rev. D}\ }\textbf {\bibinfo
  {volume} {83}},\ \bibinfo {pages} {014504} (\bibinfo {year} {2011})},\
  \Eprint {http://arxiv.org/abs/1011.3130} {arXiv:1011.3130 [hep-lat]}
  \BibitemShut {NoStop}%
\bibitem [{\citenamefont {Bonati}\ \emph {et~al.}(2015)\citenamefont {Bonati},
  \citenamefont {D'Elia}, \citenamefont {Mariti}, \citenamefont {Mesiti},
  \citenamefont {Negro},\ and\ \citenamefont {Sanfilippo}}]{Bonati:2015bha}%
  \BibitemOpen
  \bibfield  {author} {\bibinfo {author} {\bibfnamefont {C.}~\bibnamefont
  {Bonati}}, \bibinfo {author} {\bibfnamefont {M.}~\bibnamefont {D'Elia}},
  \bibinfo {author} {\bibfnamefont {M.}~\bibnamefont {Mariti}}, \bibinfo
  {author} {\bibfnamefont {M.}~\bibnamefont {Mesiti}}, \bibinfo {author}
  {\bibfnamefont {F.}~\bibnamefont {Negro}}, \ and\ \bibinfo {author}
  {\bibfnamefont {F.}~\bibnamefont {Sanfilippo}},\ }\href {\doibase
  10.1103/PhysRevD.92.054503} {\bibfield  {journal} {\bibinfo  {journal} {Phys.
  Rev. D}\ }\textbf {\bibinfo {volume} {92}},\ \bibinfo {pages} {054503}
  (\bibinfo {year} {2015})},\ \Eprint {http://arxiv.org/abs/1507.03571}
  {arXiv:1507.03571 [hep-lat]} \BibitemShut {NoStop}%
\bibitem [{\citenamefont {Bellwied}\ \emph {et~al.}(2015)\citenamefont
  {Bellwied}, \citenamefont {Borsanyi}, \citenamefont {Fodor}, \citenamefont
  {G\"unther}, \citenamefont {Katz}, \citenamefont {Ratti},\ and\ \citenamefont
  {Szabo}}]{Bellwied:2015rza}%
  \BibitemOpen
  \bibfield  {author} {\bibinfo {author} {\bibfnamefont {R.}~\bibnamefont
  {Bellwied}}, \bibinfo {author} {\bibfnamefont {S.}~\bibnamefont {Borsanyi}},
  \bibinfo {author} {\bibfnamefont {Z.}~\bibnamefont {Fodor}}, \bibinfo
  {author} {\bibfnamefont {J.}~\bibnamefont {G\"unther}}, \bibinfo {author}
  {\bibfnamefont {S.~D.}\ \bibnamefont {Katz}}, \bibinfo {author}
  {\bibfnamefont {C.}~\bibnamefont {Ratti}}, \ and\ \bibinfo {author}
  {\bibfnamefont {K.~K.}\ \bibnamefont {Szabo}},\ }\href {\doibase
  10.1016/j.physletb.2015.11.011} {\bibfield  {journal} {\bibinfo  {journal}
  {Phys. Lett. B}\ }\textbf {\bibinfo {volume} {751}},\ \bibinfo {pages} {559}
  (\bibinfo {year} {2015})},\ \Eprint {http://arxiv.org/abs/1507.07510}
  {arXiv:1507.07510 [hep-lat]} \BibitemShut {NoStop}%
\bibitem [{\citenamefont {Cea}\ \emph {et~al.}(2016)\citenamefont {Cea},
  \citenamefont {Cosmai},\ and\ \citenamefont {Papa}}]{Cea:2015cya}%
  \BibitemOpen
  \bibfield  {author} {\bibinfo {author} {\bibfnamefont {P.}~\bibnamefont
  {Cea}}, \bibinfo {author} {\bibfnamefont {L.}~\bibnamefont {Cosmai}}, \ and\
  \bibinfo {author} {\bibfnamefont {A.}~\bibnamefont {Papa}},\ }\href {\doibase
  10.1103/PhysRevD.93.014507} {\bibfield  {journal} {\bibinfo  {journal} {Phys.
  Rev. D}\ }\textbf {\bibinfo {volume} {93}},\ \bibinfo {pages} {014507}
  (\bibinfo {year} {2016})},\ \Eprint {http://arxiv.org/abs/1508.07599}
  {arXiv:1508.07599 [hep-lat]} \BibitemShut {NoStop}%
\bibitem [{\citenamefont {Bonati}\ \emph {et~al.}(2018)\citenamefont {Bonati},
  \citenamefont {D'Elia}, \citenamefont {Negro}, \citenamefont {Sanfilippo},\
  and\ \citenamefont {Zambello}}]{Bonati:2018nut}%
  \BibitemOpen
  \bibfield  {author} {\bibinfo {author} {\bibfnamefont {C.}~\bibnamefont
  {Bonati}}, \bibinfo {author} {\bibfnamefont {M.}~\bibnamefont {D'Elia}},
  \bibinfo {author} {\bibfnamefont {F.}~\bibnamefont {Negro}}, \bibinfo
  {author} {\bibfnamefont {F.}~\bibnamefont {Sanfilippo}}, \ and\ \bibinfo
  {author} {\bibfnamefont {K.}~\bibnamefont {Zambello}},\ }\href {\doibase
  10.1103/PhysRevD.98.054510} {\bibfield  {journal} {\bibinfo  {journal} {Phys.
  Rev. D}\ }\textbf {\bibinfo {volume} {98}},\ \bibinfo {pages} {054510}
  (\bibinfo {year} {2018})},\ \Eprint {http://arxiv.org/abs/1805.02960}
  {arXiv:1805.02960 [hep-lat]} \BibitemShut {NoStop}%
\bibitem [{\citenamefont {D'Elia}(2019)}]{DElia:2018fjp}%
  \BibitemOpen
  \bibfield  {author} {\bibinfo {author} {\bibfnamefont {M.}~\bibnamefont
  {D'Elia}},\ }\href {\doibase 10.1016/j.nuclphysa.2018.10.042} {\bibfield
  {journal} {\bibinfo  {journal} {Nucl. Phys. A}\ }\textbf {\bibinfo {volume}
  {982}},\ \bibinfo {pages} {99} (\bibinfo {year} {2019})},\ \Eprint
  {http://arxiv.org/abs/1809.10660} {arXiv:1809.10660 [hep-lat]} \BibitemShut
  {NoStop}%
\bibitem [{\citenamefont {Biswas}\ \emph {et~al.}(2024)\citenamefont {Biswas},
  \citenamefont {Petreczky},\ and\ \citenamefont {Sharma}}]{Biswas:2024xxh}%
  \BibitemOpen
  \bibfield  {author} {\bibinfo {author} {\bibfnamefont {D.}~\bibnamefont
  {Biswas}}, \bibinfo {author} {\bibfnamefont {P.}~\bibnamefont {Petreczky}}, \
  and\ \bibinfo {author} {\bibfnamefont {S.}~\bibnamefont {Sharma}},\
  }\href@noop {} {\  (\bibinfo {year} {2024})},\ \Eprint
  {http://arxiv.org/abs/2401.02874} {arXiv:2401.02874 [hep-ph]} \BibitemShut
  {NoStop}%
\bibitem [{\citenamefont {Clarke}\ \emph {et~al.}(2021)\citenamefont {Clarke},
  \citenamefont {Kaczmarek}, \citenamefont {Karsch}, \citenamefont {Lahiri},\
  and\ \citenamefont {Sarkar}}]{Clarke:2020htu}%
  \BibitemOpen
  \bibfield  {author} {\bibinfo {author} {\bibfnamefont {D.~A.}\ \bibnamefont
  {Clarke}}, \bibinfo {author} {\bibfnamefont {O.}~\bibnamefont {Kaczmarek}},
  \bibinfo {author} {\bibfnamefont {F.}~\bibnamefont {Karsch}}, \bibinfo
  {author} {\bibfnamefont {A.}~\bibnamefont {Lahiri}}, \ and\ \bibinfo {author}
  {\bibfnamefont {M.}~\bibnamefont {Sarkar}},\ }\href {\doibase
  10.1103/PhysRevD.103.L011501} {\bibfield  {journal} {\bibinfo  {journal}
  {Phys. Rev. D}\ }\textbf {\bibinfo {volume} {103}},\ \bibinfo {pages}
  {L011501} (\bibinfo {year} {2021})},\ \Eprint
  {http://arxiv.org/abs/2008.11678} {arXiv:2008.11678 [hep-lat]} \BibitemShut
  {NoStop}%
\bibitem [{\citenamefont {Kaczmarek}\ \emph {et~al.}(2022)\citenamefont
  {Kaczmarek}, \citenamefont {Karsch}, \citenamefont {Lahiri}, \citenamefont
  {Li}, \citenamefont {Sarkar}, \citenamefont {Schmidt},\ and\ \citenamefont
  {Scior}}]{Kaczmarek:2021ufg}%
  \BibitemOpen
  \bibfield  {author} {\bibinfo {author} {\bibfnamefont {O.}~\bibnamefont
  {Kaczmarek}}, \bibinfo {author} {\bibfnamefont {F.}~\bibnamefont {Karsch}},
  \bibinfo {author} {\bibfnamefont {A.}~\bibnamefont {Lahiri}}, \bibinfo
  {author} {\bibfnamefont {S.-T.}\ \bibnamefont {Li}}, \bibinfo {author}
  {\bibfnamefont {M.}~\bibnamefont {Sarkar}}, \bibinfo {author} {\bibfnamefont
  {C.}~\bibnamefont {Schmidt}}, \ and\ \bibinfo {author} {\bibfnamefont
  {P.}~\bibnamefont {Scior}},\ }\href {\doibase 10.22323/1.396.0429} {\bibfield
   {journal} {\bibinfo  {journal} {PoS}\ }\textbf {\bibinfo {volume}
  {LATTICE2021}},\ \bibinfo {pages} {429} (\bibinfo {year} {2022})},\ \Eprint
  {http://arxiv.org/abs/2112.15398} {arXiv:2112.15398 [hep-lat]} \BibitemShut
  {NoStop}%
\bibitem [{\citenamefont {Follana}\ \emph {et~al.}(2007)\citenamefont
  {Follana}, \citenamefont {Mason}, \citenamefont {Davies}, \citenamefont
  {Hornbostel}, \citenamefont {Lepage}, \citenamefont {Shigemitsu},
  \citenamefont {Trottier},\ and\ \citenamefont {Wong}}]{Follana:2006rc}%
  \BibitemOpen
  \bibfield  {author} {\bibinfo {author} {\bibfnamefont {E.}~\bibnamefont
  {Follana}}, \bibinfo {author} {\bibfnamefont {Q.}~\bibnamefont {Mason}},
  \bibinfo {author} {\bibfnamefont {C.}~\bibnamefont {Davies}}, \bibinfo
  {author} {\bibfnamefont {K.}~\bibnamefont {Hornbostel}}, \bibinfo {author}
  {\bibfnamefont {G.~P.}\ \bibnamefont {Lepage}}, \bibinfo {author}
  {\bibfnamefont {J.}~\bibnamefont {Shigemitsu}}, \bibinfo {author}
  {\bibfnamefont {H.}~\bibnamefont {Trottier}}, \ and\ \bibinfo {author}
  {\bibfnamefont {K.}~\bibnamefont {Wong}} (\bibinfo {collaboration} {HPQCD,
  UKQCD}),\ }\href {\doibase 10.1103/PhysRevD.75.054502} {\bibfield  {journal}
  {\bibinfo  {journal} {Phys. Rev. D}\ }\textbf {\bibinfo {volume} {75}},\
  \bibinfo {pages} {054502} (\bibinfo {year} {2007})},\ \Eprint
  {http://arxiv.org/abs/hep-lat/0610092} {arXiv:hep-lat/0610092} \BibitemShut
  {NoStop}%
\bibitem [{\citenamefont {Bazavov}\ \emph {et~al.}(2012)\citenamefont {Bazavov}
  \emph {et~al.}}]{Bazavov:2011nk}%
  \BibitemOpen
  \bibfield  {author} {\bibinfo {author} {\bibfnamefont {A.}~\bibnamefont
  {Bazavov}} \emph {et~al.},\ }\href {\doibase 10.1103/PhysRevD.85.054503}
  {\bibfield  {journal} {\bibinfo  {journal} {Phys. Rev. D}\ }\textbf {\bibinfo
  {volume} {85}},\ \bibinfo {pages} {054503} (\bibinfo {year} {2012})},\
  \Eprint {http://arxiv.org/abs/1111.1710} {arXiv:1111.1710 [hep-lat]}
  \BibitemShut {NoStop}%
\bibitem [{\citenamefont {Bollweg}\ \emph {et~al.}(2021)\citenamefont
  {Bollweg}, \citenamefont {Goswami}, \citenamefont {Kaczmarek}, \citenamefont
  {Karsch}, \citenamefont {Mukherjee}, \citenamefont {Petreczky}, \citenamefont
  {Schmidt},\ and\ \citenamefont {Scior}}]{Bollweg:2021vqf}%
  \BibitemOpen
  \bibfield  {author} {\bibinfo {author} {\bibfnamefont {D.}~\bibnamefont
  {Bollweg}}, \bibinfo {author} {\bibfnamefont {J.}~\bibnamefont {Goswami}},
  \bibinfo {author} {\bibfnamefont {O.}~\bibnamefont {Kaczmarek}}, \bibinfo
  {author} {\bibfnamefont {F.}~\bibnamefont {Karsch}}, \bibinfo {author}
  {\bibfnamefont {S.}~\bibnamefont {Mukherjee}}, \bibinfo {author}
  {\bibfnamefont {P.}~\bibnamefont {Petreczky}}, \bibinfo {author}
  {\bibfnamefont {C.}~\bibnamefont {Schmidt}}, \ and\ \bibinfo {author}
  {\bibfnamefont {P.}~\bibnamefont {Scior}} (\bibinfo {collaboration}
  {HotQCD}),\ }\href {\doibase 10.1103/PhysRevD.104.074512} {\bibfield
  {journal} {\bibinfo  {journal} {Phys. Rev. D}\ }\textbf {\bibinfo {volume}
  {104}} (\bibinfo {year} {2021}),\ 10.1103/PhysRevD.104.074512},\ \Eprint
  {http://arxiv.org/abs/2107.10011} {arXiv:2107.10011 [hep-lat]} \BibitemShut
  {NoStop}%
\bibitem [{\citenamefont {Aoki}\ \emph {et~al.}(2020)\citenamefont {Aoki} \emph
  {et~al.}}]{FlavourLatticeAveragingGroup:2019iem}%
  \BibitemOpen
  \bibfield  {author} {\bibinfo {author} {\bibfnamefont {S.}~\bibnamefont
  {Aoki}} \emph {et~al.} (\bibinfo {collaboration} {Flavour Lattice Averaging
  Group}),\ }\href {\doibase 10.1140/epjc/s10052-019-7354-7} {\bibfield
  {journal} {\bibinfo  {journal} {Eur. Phys. J. C}\ }\textbf {\bibinfo {volume}
  {80}},\ \bibinfo {pages} {113} (\bibinfo {year} {2020})},\ \Eprint
  {http://arxiv.org/abs/1902.08191} {arXiv:1902.08191 [hep-lat]} \BibitemShut
  {NoStop}%
\bibitem [{\citenamefont {Bollweg}\ \emph {et~al.}(2022)\citenamefont
  {Bollweg}, \citenamefont {Altenkort}, \citenamefont {Clarke}, \citenamefont
  {Kaczmarek}, \citenamefont {Mazur}, \citenamefont {Schmidt}, \citenamefont
  {Scior},\ and\ \citenamefont {Shu}}]{Bollweg:2021cvl}%
  \BibitemOpen
  \bibfield  {author} {\bibinfo {author} {\bibfnamefont {D.}~\bibnamefont
  {Bollweg}}, \bibinfo {author} {\bibfnamefont {L.}~\bibnamefont {Altenkort}},
  \bibinfo {author} {\bibfnamefont {D.~A.}\ \bibnamefont {Clarke}}, \bibinfo
  {author} {\bibfnamefont {O.}~\bibnamefont {Kaczmarek}}, \bibinfo {author}
  {\bibfnamefont {L.}~\bibnamefont {Mazur}}, \bibinfo {author} {\bibfnamefont
  {C.}~\bibnamefont {Schmidt}}, \bibinfo {author} {\bibfnamefont
  {P.}~\bibnamefont {Scior}}, \ and\ \bibinfo {author} {\bibfnamefont {H.-T.}\
  \bibnamefont {Shu}},\ }\href {\doibase 10.22323/1.396.0196} {\bibfield
  {journal} {\bibinfo  {journal} {PoS}\ }\textbf {\bibinfo {volume}
  {LATTICE2021}},\ \bibinfo {pages} {196} (\bibinfo {year} {2022})},\ \Eprint
  {http://arxiv.org/abs/2111.10354} {arXiv:2111.10354 [hep-lat]} \BibitemShut
  {NoStop}%
\bibitem [{\citenamefont {Mazur}\ \emph {et~al.}(2023)\citenamefont {Mazur}
  \emph {et~al.}}]{HotQCD:2023ghu}%
  \BibitemOpen
  \bibfield  {author} {\bibinfo {author} {\bibfnamefont {L.}~\bibnamefont
  {Mazur}} \emph {et~al.} (\bibinfo {collaboration} {HotQCD}),\ }\href@noop {}
  {\  (\bibinfo {year} {2023})},\ \Eprint {http://arxiv.org/abs/2306.01098}
  {arXiv:2306.01098 [hep-lat]} \BibitemShut {NoStop}%
\bibitem [{\citenamefont {Gavai}\ and\ \citenamefont
  {Gupta}(2003)}]{Gavai:2003mf}%
  \BibitemOpen
  \bibfield  {author} {\bibinfo {author} {\bibfnamefont {R.~V.}\ \bibnamefont
  {Gavai}}\ and\ \bibinfo {author} {\bibfnamefont {S.}~\bibnamefont {Gupta}},\
  }\href {\doibase 10.1103/PhysRevD.68.034506} {\bibfield  {journal} {\bibinfo
  {journal} {Phys. Rev. D}\ }\textbf {\bibinfo {volume} {68}},\ \bibinfo
  {pages} {034506} (\bibinfo {year} {2003})},\ \Eprint
  {http://arxiv.org/abs/hep-lat/0303013} {arXiv:hep-lat/0303013} \BibitemShut
  {NoStop}%
\bibitem [{\citenamefont {Allton}\ \emph {et~al.}(2005)\citenamefont {Allton},
  \citenamefont {Doring}, \citenamefont {Ejiri}, \citenamefont {Hands},
  \citenamefont {Kaczmarek}, \citenamefont {Karsch}, \citenamefont {Laermann},\
  and\ \citenamefont {Redlich}}]{Allton:2005gk}%
  \BibitemOpen
  \bibfield  {author} {\bibinfo {author} {\bibfnamefont {C.~R.}\ \bibnamefont
  {Allton}}, \bibinfo {author} {\bibfnamefont {M.}~\bibnamefont {Doring}},
  \bibinfo {author} {\bibfnamefont {S.}~\bibnamefont {Ejiri}}, \bibinfo
  {author} {\bibfnamefont {S.~J.}\ \bibnamefont {Hands}}, \bibinfo {author}
  {\bibfnamefont {O.}~\bibnamefont {Kaczmarek}}, \bibinfo {author}
  {\bibfnamefont {F.}~\bibnamefont {Karsch}}, \bibinfo {author} {\bibfnamefont
  {E.}~\bibnamefont {Laermann}}, \ and\ \bibinfo {author} {\bibfnamefont
  {K.}~\bibnamefont {Redlich}},\ }\href {\doibase 10.1103/PhysRevD.71.054508}
  {\bibfield  {journal} {\bibinfo  {journal} {Phys. Rev. D}\ }\textbf {\bibinfo
  {volume} {71}},\ \bibinfo {pages} {054508} (\bibinfo {year} {2005})},\
  \Eprint {http://arxiv.org/abs/hep-lat/0501030} {arXiv:hep-lat/0501030}
  \BibitemShut {NoStop}%
\bibitem [{\citenamefont {Cheng}\ \emph {et~al.}(2008)\citenamefont {Cheng}
  \emph {et~al.}}]{Cheng:2007jq}%
  \BibitemOpen
  \bibfield  {author} {\bibinfo {author} {\bibfnamefont {M.}~\bibnamefont
  {Cheng}} \emph {et~al.},\ }\href {\doibase 10.1103/PhysRevD.77.014511}
  {\bibfield  {journal} {\bibinfo  {journal} {Phys. Rev. D}\ }\textbf {\bibinfo
  {volume} {77}},\ \bibinfo {pages} {014511} (\bibinfo {year} {2008})},\
  \Eprint {http://arxiv.org/abs/0710.0354} {arXiv:0710.0354 [hep-lat]}
  \BibitemShut {NoStop}%
\bibitem [{\citenamefont {Ejiri}\ \emph {et~al.}(2009)\citenamefont {Ejiri},
  \citenamefont {Karsch}, \citenamefont {Laermann}, \citenamefont {Miao},
  \citenamefont {Mukherjee}, \citenamefont {Petreczky}, \citenamefont
  {Schmidt}, \citenamefont {Soeldner},\ and\ \citenamefont
  {Unger}}]{Ejiri:2009ac}%
  \BibitemOpen
  \bibfield  {author} {\bibinfo {author} {\bibfnamefont {S.}~\bibnamefont
  {Ejiri}}, \bibinfo {author} {\bibfnamefont {F.}~\bibnamefont {Karsch}},
  \bibinfo {author} {\bibfnamefont {E.}~\bibnamefont {Laermann}}, \bibinfo
  {author} {\bibfnamefont {C.}~\bibnamefont {Miao}}, \bibinfo {author}
  {\bibfnamefont {S.}~\bibnamefont {Mukherjee}}, \bibinfo {author}
  {\bibfnamefont {P.}~\bibnamefont {Petreczky}}, \bibinfo {author}
  {\bibfnamefont {C.}~\bibnamefont {Schmidt}}, \bibinfo {author} {\bibfnamefont
  {W.}~\bibnamefont {Soeldner}}, \ and\ \bibinfo {author} {\bibfnamefont
  {W.}~\bibnamefont {Unger}},\ }\href {\doibase 10.1103/PhysRevD.80.094505}
  {\bibfield  {journal} {\bibinfo  {journal} {Phys. Rev. D}\ }\textbf {\bibinfo
  {volume} {80}},\ \bibinfo {pages} {094505} (\bibinfo {year} {2009})},\
  \Eprint {http://arxiv.org/abs/0909.5122} {arXiv:0909.5122 [hep-lat]}
  \BibitemShut {NoStop}%
\bibitem [{\citenamefont {Unger}(2010)}]{Unger:2010wcq}%
  \BibitemOpen
  \bibfield  {author} {\bibinfo {author} {\bibfnamefont {W.}~\bibnamefont
  {Unger}},\ }\emph {\bibinfo {title} {{The chiral phase transition of QCD with
  2+1 flavors : a lattice study on Goldstone modes and universal scaling}}},\
  \href@noop {} {Ph.D. thesis},\ \bibinfo  {school} {U. Bielefeld} (\bibinfo
  {year} {2010}),\ \bibinfo {note}
  {\url{https://pub.uni-bielefeld.de/record/2302381}}\BibitemShut {NoStop}%
\bibitem [{\citenamefont {Engels}\ \emph {et~al.}(2000)\citenamefont {Engels},
  \citenamefont {Holtmann}, \citenamefont {Mendes},\ and\ \citenamefont
  {Schulze}}]{Engels:2000xw}%
  \BibitemOpen
  \bibfield  {author} {\bibinfo {author} {\bibfnamefont {J.}~\bibnamefont
  {Engels}}, \bibinfo {author} {\bibfnamefont {S.}~\bibnamefont {Holtmann}},
  \bibinfo {author} {\bibfnamefont {T.}~\bibnamefont {Mendes}}, \ and\ \bibinfo
  {author} {\bibfnamefont {T.}~\bibnamefont {Schulze}},\ }\href {\doibase
  10.1016/S0370-2693(00)01079-0} {\bibfield  {journal} {\bibinfo  {journal}
  {Phys. Lett. B}\ }\textbf {\bibinfo {volume} {492}},\ \bibinfo {pages} {219}
  (\bibinfo {year} {2000})},\ \Eprint {http://arxiv.org/abs/hep-lat/0006023}
  {arXiv:hep-lat/0006023} \BibitemShut {NoStop}%
\bibitem [{\citenamefont {Hasenbusch}\ and\ \citenamefont
  {Toeroek}(1999)}]{Hasenbusch:1999cc}%
  \BibitemOpen
  \bibfield  {author} {\bibinfo {author} {\bibfnamefont {M.}~\bibnamefont
  {Hasenbusch}}\ and\ \bibinfo {author} {\bibfnamefont {T.}~\bibnamefont
  {Toeroek}},\ }\href {\doibase 10.1088/0305-4470/32/36/301} {\bibfield
  {journal} {\bibinfo  {journal} {J. Phys. A}\ }\textbf {\bibinfo {volume}
  {32}},\ \bibinfo {pages} {6361} (\bibinfo {year} {1999})},\ \Eprint
  {http://arxiv.org/abs/cond-mat/9904408} {arXiv:cond-mat/9904408} \BibitemShut
  {NoStop}%
\bibitem [{\citenamefont {Karsch}\ \emph {et~al.}(2023)\citenamefont {Karsch},
  \citenamefont {Neumann},\ and\ \citenamefont {Sarkar}}]{Karsch:2023pga}%
  \BibitemOpen
  \bibfield  {author} {\bibinfo {author} {\bibfnamefont {F.}~\bibnamefont
  {Karsch}}, \bibinfo {author} {\bibfnamefont {M.}~\bibnamefont {Neumann}}, \
  and\ \bibinfo {author} {\bibfnamefont {M.}~\bibnamefont {Sarkar}},\ }\href
  {\doibase 10.1103/PhysRevD.108.014505} {\bibfield  {journal} {\bibinfo
  {journal} {Phys. Rev. D}\ }\textbf {\bibinfo {volume} {108}},\ \bibinfo
  {pages} {014505} (\bibinfo {year} {2023})},\ \Eprint
  {http://arxiv.org/abs/2304.01710} {arXiv:2304.01710 [hep-lat]} \BibitemShut
  {NoStop}%
\bibitem [{\citenamefont {Blaizot}\ \emph {et~al.}(2001)\citenamefont
  {Blaizot}, \citenamefont {Iancu},\ and\ \citenamefont
  {Rebhan}}]{Blaizot:2001vr}%
  \BibitemOpen
  \bibfield  {author} {\bibinfo {author} {\bibfnamefont {J.~P.}\ \bibnamefont
  {Blaizot}}, \bibinfo {author} {\bibfnamefont {E.}~\bibnamefont {Iancu}}, \
  and\ \bibinfo {author} {\bibfnamefont {A.}~\bibnamefont {Rebhan}},\ }\href
  {\doibase 10.1016/S0370-2693(01)01316-8} {\bibfield  {journal} {\bibinfo
  {journal} {Phys. Lett. B}\ }\textbf {\bibinfo {volume} {523}},\ \bibinfo
  {pages} {143} (\bibinfo {year} {2001})},\ \Eprint
  {http://arxiv.org/abs/hep-ph/0110369} {arXiv:hep-ph/0110369} \BibitemShut
  {NoStop}%
\bibitem [{\citenamefont {Bazavov}\ \emph {et~al.}(2017)\citenamefont {Bazavov}
  \emph {et~al.}}]{Bazavov:2017dus}%
  \BibitemOpen
  \bibfield  {author} {\bibinfo {author} {\bibfnamefont {A.}~\bibnamefont
  {Bazavov}} \emph {et~al.},\ }\href {\doibase 10.1103/PhysRevD.95.054504}
  {\bibfield  {journal} {\bibinfo  {journal} {Phys. Rev. D}\ }\textbf {\bibinfo
  {volume} {95}},\ \bibinfo {pages} {054504} (\bibinfo {year} {2017})},\
  \Eprint {http://arxiv.org/abs/1701.04325} {arXiv:1701.04325 [hep-lat]}
  \BibitemShut {NoStop}%
\bibitem [{\citenamefont {Ding}\ \emph {et~al.}(2024)\citenamefont {Ding},
  \citenamefont {Kaczmarek}, \citenamefont {Karsch}, \citenamefont {Petreczky},
  \citenamefont {Sarkar}, \citenamefont {Schmidt},\ and\ \citenamefont
  {Sharma}}]{ding2024dataset}%
  \BibitemOpen
  \bibfield  {author} {\bibinfo {author} {\bibfnamefont {H.-T.}\ \bibnamefont
  {Ding}}, \bibinfo {author} {\bibfnamefont {O.}~\bibnamefont {Kaczmarek}},
  \bibinfo {author} {\bibfnamefont {F.}~\bibnamefont {Karsch}}, \bibinfo
  {author} {\bibfnamefont {P.}~\bibnamefont {Petreczky}}, \bibinfo {author}
  {\bibfnamefont {M.}~\bibnamefont {Sarkar}}, \bibinfo {author} {\bibfnamefont
  {C.}~\bibnamefont {Schmidt}}, \ and\ \bibinfo {author} {\bibfnamefont
  {S.}~\bibnamefont {Sharma}},\ }\href {\doibase 10.4119/unibi/2990193}
  {\bibfield  {journal} {\bibinfo  {journal} {Bielefeld University}\ }
  (\bibinfo {year} {2024}),\ 10.4119/unibi/2990193}\BibitemShut {NoStop}%
\bibitem [{\citenamefont {Ali}\ \emph {et~al.}(2024)\citenamefont {Ali},
  \citenamefont {Biswas}, \citenamefont {Jaiswal},\ and\ \citenamefont
  {Mishra}}]{Ali:2024nrz}%
  \BibitemOpen
  \bibfield  {author} {\bibinfo {author} {\bibfnamefont {M.~S.}\ \bibnamefont
  {Ali}}, \bibinfo {author} {\bibfnamefont {D.}~\bibnamefont {Biswas}},
  \bibinfo {author} {\bibfnamefont {A.}~\bibnamefont {Jaiswal}}, \ and\
  \bibinfo {author} {\bibfnamefont {H.}~\bibnamefont {Mishra}},\ }\href
  {\doibase 10.1103/PhysRevD.109.114017} {\bibfield  {journal} {\bibinfo
  {journal} {Phys. Rev. D}\ }\textbf {\bibinfo {volume} {109}},\ \bibinfo
  {pages} {114017} (\bibinfo {year} {2024})},\ \Eprint
  {http://arxiv.org/abs/2403.11965} {arXiv:2403.11965 [nucl-th]} \BibitemShut
  {NoStop}%
\end{thebibliography}

\end{document}